\documentclass[11pt]{article}
\usepackage[margin=1in]{geometry}
\usepackage{graphicx} 
\usepackage[utf8]{inputenc}
\usepackage[font=small,labelfont=bf]{caption}
\usepackage{graphicx}
\usepackage{float}
\usepackage{subcaption}
\usepackage[numbers,sort&compress,elide]{natbib}
\usepackage{bm}
\usepackage{ulem}
\usepackage{hyperref}
\usepackage{authblk}  
\usepackage{siunitx}
\DeclareSIUnit\bar{bar}
\usepackage{mathtools}
\usepackage{array}

\hypersetup{colorlinks=true,citecolor=blue, linkcolor=blue,urlcolor=blue}
\graphicspath{{figures/}}   

\begin{document}
\title{Design of high-efficiency UHV loading of nanodiamonds into a Paul trap: Towards Matter-Wave Interferometry with Massive Objects}
\author[1]{Rafael Benjaminov}
\author[1]{Sela Liran}
\author[1]{Or Dobkowski}
\author[1]{Yaniv Bar-Haim}
\author[1]{Michael Averbukh}
\author[1]{Ron Folman}

\affil[1]{Ben-Gurion University of the Negev, Department of Physics and Ilse Katz Institute for Nanoscale Science and Technology, Be'er Sheva 84105, Israel}
\maketitle

\begin{abstract}
Quantum mechanics (QM) and General relativity (GR), also known as the theory of gravity, are the two pillars of modern physics. A matter-wave interferometer with a massive particle, can test numerous fundamental ideas, including the spatial superposition principle - a foundational concept in QM - in completely new regimes, as well as the interface between QM and GR, e.g., testing the quantization of gravity. Consequently, there exists an intensive effort to realize such an interferometer. While several paths are being pursued, we focus on utilizing nanodiamonds as our particle, and a spin embedded in the ND together with Stern-Gerlach forces, to achieve a closed loop in space-time. There is a growing community of groups pursuing this path \cite{whitepaper_cern_xxx}. We are posting this technical note (as part of a series of seven such notes), to highlight our plans and solutions concerning various challenges in this ambitious endeavor, hoping this will support this growing community. In this work, we review current methods for loading nanodiamonds into a Paul trap, and their capabilities and limitations regarding our application. We also present our experiments on loading and launching nanodiamonds using a vibrating piezoelectric element and by electrical forces. Finally, we present our design of a novel nanodiamond loading method for ultra-high-vacuum experiments. As the production of highly accurate, high-purity nanodiamonds with a single NV required for interferometric measurements is expected to be expensive, we put emphasis on achieving high loading efficiency, while loading the charged ND into a Paul trap in ultra-high vacuum.
\end{abstract}

\section{Introduction}
\subsection{Nanodiamond interferometer as a probe of quantum gravity}
Quantum mechanics (QM) is a pillar of modern physics. It is thus imperative to test it in ever-growing regions of the relevant parameter space. A second pillar is general relativity (GR), and as a unification of the two seems to be eluding continuous theoretical efforts, it is just as imperative to experimentally test the interface of these two pillars by conducting experiments in which foundational concepts of the two theories must work in concert.

The most advanced demonstrations of massive spatial superpositions have been achieved by Markus Arndt's group, reaching systems composed of approximately 2,000 atoms \cite{Fein2019MoleculeSuperpositions}. This will surely grow by one or two orders of magnitude in the near future. An important question is whether one can find a new technique that would push the state of the art much further in the mass and spatial extent of the superposition. Several paths are being pursued \cite{RomeroIsart2017CoherentInflation, Pino2018OnChipSuperconductingMicrosphere, Weiss2021LargeDelocalizationOptimalControl, Neumeier2024FastQuantumInterference, Kialka2022RoadmapHighMassMWI}, and we choose to utilize Stern-Gerlach forces.

The Stern-Gerlach interferometer (SGI) has, in the last decade, proven to be an agile tool for atom interferometry \cite{Amit2019T3SGInterferometer,Dobkowski2025QuantumEquivalence, Keil2021SternGerlachAtomChip}. Consequently, we, as well as others, aim to utilize it for interferometry with massive particles, specifically, nanodiamonds (NDs) with a single spin embedded in their center \cite{Wan2016FreeRamsey,Scala2013SpinInducedMWI,Margalit2021CompleteSGI}.

Levitating, trapping, and cooling of massive particles, most probably a prerequisite for interferometry with such particles, has made significant progress in recent years. Specifically, the field of levitodynamics is a fast-growing field \cite{gonzalez-ballestero_levitodynamics_2021}. Commonly used particles are silica spheres. As the state of the art spans a wide spectrum of techniques, achievements and applications, instead of referencing numerous works, we take, for the benefit of the reader, the unconventional step of simply mapping some of the principal investigators; these include Markus Aspelmeyer, Lukas Novotny, Peter Barker, Kiyotaka Aikawa, Romain Quidant, Francesco Marin, Hendrik Ulbricht and David Moore. Relevant to this work, a rather new sub-field which is now being developed deals with ND particles, where the significant difference is that a spin with long coherence times may be embedded in the ND. Such a spin, originating from a nitrogen-vacancy (NV) center, could enable the coherent splitting and recombination of the ND by utilizing Stern-Gerlach forces \cite{Margalit2021CompleteSGI, Wan2016FreeRamsey, Scala2013SpinInducedMWI}. This endeavor includes principal investigators
such as Tongcang Li, Gavin Morley, Gabriel Hetet, Tracy Northup, Brian D’Urso, Andrew Geraci, Jason Twamley and Gurudev Dutt.

In this work, we aim to start with a ND of $10^7$ atoms and extremely short interferometer durations. Closing a loop in space-time in a very short time is enabled by the strong magnetic gradients induced by the current-carrying wires of the atom chip \cite{Keil2016FifteenYearsAtomChip}. Such an interferometer will already enable testing the existing understanding concerning environmental decoherence (e.g., from blackbody radiation), and internal decoherence \cite{HenkelFolman2024UniversalLimitPhonons}, never tested on such a large object in a spatial superposition. As we slowly evolve to higher masses and longer durations (larger splitting), the ND SGI will enable the community to probe not only the superposition principle in completely new regimes, but in addition, it will enable us to test specific aspects of exotic ideas such as the Continuous spontaneous localization hypothesis \cite{Adler2021CSLLayering,Gasbarri2021TestingFoundationsSpace}. As the masses are increased, the ND SGI will be able to test hypotheses related to gravity, such as modifications to gravity at short ranges (also known as the fifth force), as one of the SGI paths may be brought in a controlled manner extremely close to a massive object \cite{Geraci2010ShortRangeForceMSpheres,GeraciGoldman2015ShortRangeNanosphereMWI,Bobowski2024ShortRangeAnisotropic,Panda2024LatticeGravAttraction}. Once SGI technology allows for even larger masses ($10^{11}$ atoms), we could test the Diósi–Penrose collapse hypothesis \cite{Penrose2014GravitizationQM,FuentesPenrose2018QuantumStateReductionBEC,Howl2019BECUnification,Tomaz2024CollapseTimeMolecules,Bassi2013CollapseModelsRMP} and gravity self-interaction \cite{HatifiDurt2023HumptyDumpty,Grossardt2021DephasingSemiclassical,AguiarMatsas2024SchrodingerNewtonSG} (e.g., the Schrödinger-Newton equation). Here starts the regime of active masses, whereby not only the gravitation of Earth needs to be taken into account. Furthermore, it is claimed that placing two such SGIs in parallel will allow probing the quantum nature of gravity \cite{Bose2017SpinEntanglementQG,MarlettoVedral2017GravInducedEntanglement}. This will be enabled by ND SGI, as with $10^{11}$ atoms, the gravitational interaction could be the strongest \cite{VanDeKamp2020CasimirScreening,Schut2023RelaxationQGIM,Schut2024MicronSizeQGEM}.

Let us emphasize that, although high accelerations may be obtained with multiple spins, we consider only an ND with a single spin, as numerous spins will result in multiple trajectories and will smear the interferometer signal. We also note that working with a ND with less than $10^7$ atoms is probably not feasible because of two reasons. The first is that NVs that are closer to the surface than 20\,nm lose coherence, and the second is that at sizes smaller than 50\,nm, the relative fabrication errors become large, and a high-precision ND source becomes beyond reach.

Although presently available NDs and ND sources are good enough for the first generation of ND SGI \cite{Feldman_Paul_trap_ND, Skakunenko_Needle_trap}, we are working on improving the quality of the NDs \cite{givon2025fabricationnanodiamondssinglenv} as well as the source, while in parallel working to realize the first generation of a ND SGI with the available technology. Here we present our experiments with existing technologies and our design of a new ND source for Paul-trap levitation in ultra-high vacuum (UHV).

This technical note is part of a series of seven technical notes put on the archive towards the end of August 2025,
including a wide range of required building blocks for a ND SGI\,\cite{Feldman_Paul_trap_ND, Skakunenko_Needle_trap, Levi_Quantum_control_NV, muretova_parametric_2025, Liran_ND_neutralization, givon2025fabricationnanodiamondssinglenv}.

\subsection{Loading nanoparticles into levitation traps}
Matter-wave interferometry with nanoparticles demands robust trapping, efficient particle loading, and ultra-high-vacuum (UHV) conditions. Efficient particle loading ensures the efficiency, reliability and repetition rate of the experiment, while robust trapping decreases uncertainty in the particle's position and allows active cooling. Achieving UHV is critical for minimizing unwanted interaction of the particles with the residual gas, thereby extending the coherence time.

Another important requirement for the ND SGI is the quality and features of the trapped ND, namely, the accurate size and shape of the ND, the position of the NV center embedded in the ND, and the purity of the diamond crystal. We are in the process of developing efficient methods to produce such custom and accurate NDs \cite{givon2025fabricationnanodiamondssinglenv}. The NDs will be nanofabricated in a cuboid shape to allow cooling of all three rotational degrees of freedom, and an NV center will be deterministically embedded in the middle of the ND. The NDs will be fabricated from a high-purity diamond to improve the spin coherence time of the NV center. The production of such highly accurate, high-purity NDs with a single NV required for interferometric measurements is expected to be expensive, even after the fabrication process R\&D has been finalized; thus, this work is motivated by the need to achieve high loading efficiency. In summary, the loading technique must maintain UHV conditions in the main science chamber, and also have high efficiency in terms of the number of trapped nanoparticles vs. the number of nanoparticles wasted in the loading process.

Two common techniques for loading nanoparticles into traps are nebulizers and electrosprays, in which the nanoparticles are suspended in a solution that is emitted towards the trap as aerosol. These methods are attractive for their simplicity and reliability, yet their reliance on liquid suspensions critically limits the minimum achievable pressure and the time required to reach it. Hence, we aim to introduce a dry loading method. Another limiting factor of these methods is their efficiency (defined as the ratio of trapped particles to total particles used), which is on the order of $10^{-10}$, which again renders them inappropriate for our goal of high efficiency loading. Promising dry techniques demonstrated in recent years include:
\begin{itemize}
    \item Laser-Induced Acoustic Desorption (LIAD)\cite{nikkhou_direct_2021,bykov_direct_2019}
    \item Hollow Core Fiber Loading (HCF) \cite{lindner_hollow-core_2024}
    \item Electric Forces Related Loading\cite{delord_electron_2017}
    \item Resonant Vibrating Piezoelectric Transducer (PZT)\cite{Almuqhim_development_2019,Atherton_sensitive_2015,Jerard_Vincent_Rubio_Ang_development,fonseca_nonlinear_2016,khodaee_dry_2022,weisman_apparatus_2022}

\end{itemize}

In the following, we review these methods and present a summary in Table \ref{tab:loading_trapping}. Laser-Induced Acoustic Desorption (LIAD) launches nanoparticles by irradiating the rear face of a thin foil with a short laser pulse. The ensuing acoustic shockwave ejects surface-bound particles toward the trap in a dry manner. Initial experiments demonstrated that commercial silica spheres with diameters of roughly \SI{100}{\nano\metre} could be injected directly into optical traps at pressures of a few mbar with high particle flux and minimal contamination \cite{nikkhou_direct_2021}. Later studies extended the technique to UHV conditions by synchronizing fast voltage switching of a Paul trap with the timing of the LIAD pulse, enabling reliable capture at background pressures as low as \SI{4e-7}{\milli\bar} while retaining high success rate of $60\,\%$\,\cite{bykov_direct_2019}. Although this technique introduces a UHV-compatible efficient source for loading nanoparticles into a Paul trap, it also introduces heating of the foil (and perhaps also the particles) due to the laser pulses, which contradicts our goal of avoiding heating of the chamber and the particles, as heating of the ND will reduce the coherence time, and heating the chamber will increase decoherence due to black body radiation (BBR).

HCF loading uses a hollow-core photonic crystal fiber to transfer a particle from a "loading" chamber into a UHV chamber, enabling deterministic, contamination-free delivery of nanoparticles. In the loading chamber, nanoparticles are introduced into the HCF and captured by a standing-wave “optical conveyor belt” formed by two counter-propagating 1064 nm beams. By detuning the beams, the particles are moved along the fiber. The fiber used in this work \cite{lindner_hollow-core_2024} has a core diameter of $9\,\rm \mu m$ and a length of 1.4\,m acting as a differential pumping tube and sustaining a differential pressure up $\Delta P = 10^{-12}\,\mathrm{mbar}$. As a result, particles travel from the higher-pressure loading side to the UHV side without compromising the UHV conditions. At the fiber’s UHV exit, the particles are trapped in the target trap. The HCF tip can be positioned with micron precision, allowing site-selective loading, retrieval, and re-deposition into chosen antinodes of a standing-wave trap. Transfer of 100-755\,nm silica particles via HCF into a vacuum chamber and trapping at a pressure of \SI{e-9}{\milli\bar} was demonstrated \cite{lindner_hollow-core_2024}. This method shows a few merits, mainly the ability to maintain UHV conditions in the main science chamber. However, it has two limitations regarding our purpose of ND matter-wave interferometry. First is the use of laser beams for transferring the particles through the HCF. NDs typically heat up from the interaction with light, reducing the NV spin coherence time, and may even graphitize and be destroyed by the interaction with the laser. Second, the particles were introduced to the HCF via an ultrasonic nebulizer; thus, the efficiency is low.

Electric launching and trapping was observed in ambient conditions, where micron and sub-micron-sized diamond particles were introduced in the vicinity of an active Paul trap and were pulled from a copper wire into the trap \cite{delord_electron_2017}. The acting forces that allow the launching towards the trap are not clear. The authors suggest several mechanisms, for example, the ponderomotive force of the Paul trap acting on the electrons in the wire or on the NDs. To understand the exact mechanism, an in-depth study is required. In this work, we demonstrate high-efficiency electric launching by operating a ring Paul trap near a conducting surface on which NDs are placed. We observe that the launching and trapping are correlated with the electric breakdown of air near the trap, which implies that the breakdown is essential to this kind of launching. The details of our experiments with this method are given in Sec. \ref{section:Results}. Our preliminary results also suggest that such a system can trap NDs with high efficiency, larger than $10^{-2}$, as we were able to load particles into the trap from a surface containing only a few tens of NDs. Electrical launching (without trapping) was also demonstrated in a series of works on an "Electrodynamic Shield" system designed by NASA, to protect and remove dust from sensitive equipment such as lenses and solar panels \cite{patel_comprehensive_2023}, in which a set of electrodes produces electric fields that can repel micron-sized dust particles from surfaces.

The piezo-driven launch technique typically employs a ring-shaped piezoelectric transducer bonded to a glass slide that holds the nanoparticles. The glass slide is used due to its low surface energy and can be coated with ultra-hydrophobic layers, which decrease even further the surface energy. When driven at its mechanical resonance with a high-voltage, high-frequency signal, the piezo generates out-of-plane vibrations that dislodge the particles. Loading of nanoparticles into a dipole trap was demonstrated with silica and polystyrene particles as small as 50\,nm \cite{khodaee_dry_2022,Jerard_Vincent_Rubio_Ang_development}. Piezo-driven loading into a Paul trap was demonstrated for $10\,\rm \mu m$ silica particles in ambient conditions \cite{Almuqhim_development_2019}. In this work, several difficulties are mentioned, such as launching smaller particles was made possible only by submerging them in a liquid, insufficient charging of the particles, and electrical breakdown between the trap and the piezo. An attempt to introduce promising UHV compatibility of this method has been conducted in later works, which were able to load silica particles of 100-300\,nm to a dipole trap or a Paul trap in sub mbar pressure \cite{geraci_sensing_2015,weisman_apparatus_2022,fonseca_nonlinear_2016}, but the method was not successfully extended to high vacuum. There are several serious drawbacks to this method. First is the high electric power required to operate the piezo, which requires high voltage of almost 100\,V, large drive currents in the 100-500\,mA order, and hundreds of kilohertz drive frequencies to achieve the required vibration amplitude, which imposes significant complexity on the driving electronics and thermal management of the piezo element. Such heating is unfavorable, as mentioned earlier, as it can hinder the coherence time of the particles and increase BBR. Second, it seems that launching particles upwards toward a trap becomes hard or impossible as the size of the particle reduces and reaches tens of nanometers, as the works that trapped particles in this size range did not launch upwards but rather dropped the nanoparticles downwards or sideways (see Table \ref{tab:loading_trapping}). Third is the lack of a systematic charging method, which is relevant for loading into the Paul trap. In a work that used piezo loading into a Paul trap, it is mentioned that the particle had a total charge of only a few elementary charges (1-3 elementary charges) \cite{fonseca_nonlinear_2016}. Piezo-driven launching is also inaccurate and has a wide scattering angle, which can introduce nanoparticle dust into the chamber.
\newcommand{\cwA}{1.5cm}  
\newcommand{\cwB}{1.7cm}  
\newcommand{\cwC}{1.6cm}  
\newcommand{\cwD}{1.4cm}  
\newcommand{\cwE}{1.4cm}  
\newcommand{\cwF}{3.6cm}  
\newcommand{\cwG}{3cm}    

\begin{table}[htbp]
\centering
\footnotesize
\setlength{\tabcolsep}{3pt}
\renewcommand{\arraystretch}{1.15}
\begin{tabular}{|>{\centering\arraybackslash}m{\cwA}|
                >{\centering\arraybackslash}m{\cwB}|
                >{\centering\arraybackslash}m{\cwC}|
                >{\centering\arraybackslash}m{\cwD}|
                >{\centering\arraybackslash}m{\cwE}|
                >{\centering\arraybackslash}m{\cwF}|
                >{\centering\arraybackslash}m{\cwG}|}
\hline
\rule{0pt}{4.2ex}\shortstack[c]{\textbf{Author--}\\\textbf{Year}\\\textbf{(Group)}} &
\rule{0pt}{4.2ex}\shortstack[c]{\textbf{Loading}\\\textbf{Method}} &
\rule{0pt}{4.2ex}\shortstack[c]{\textbf{Particle's}\\\textbf{Type and}\\\textbf{Size}} &
\rule{0pt}{4.2ex}\shortstack[c]{\textbf{Trapping}\\\textbf{Method}} &
\rule{0pt}{4.2ex}\shortstack[c]{\textbf{Pressure}\\\textbf{[mbar]}} &
\rule{0pt}{4.2ex}\textbf{Details} &
\rule{0pt}{4.2ex}\textbf{Limitations} \\
\hline
Atherton-2015 (Geraci) \cite{Atherton_sensitive_2015} & PZT & \shortstack[c]{Silica\\300\,nm} & Dipole trap & 0.6 &
Particles are put on the lower surface of glass (clamped to the piezo) and fall due to piezo vibrations. &
Low flux of particles. Hard trapping at vacuum without cooling. Decreasing trapping efficiency with decreasing particle size.\\ \hline
Fonseca-2016 (Barker) \cite{fonseca_nonlinear_2016} & PZT & \shortstack[c]{Silica\\209\,nm} & Hybrid Paul trap & 0.5 &
Particles are introduced into the hybrid trap by initially placing them on a vibrating piezo. &
Not enough charge for stable Paul trapping at vacuum. \\ \hline
Almuqhim-2019 (Barker) \cite{Almuqhim_development_2019} & PZT & \shortstack[c]{Silica\\10 \(\mu\)m} & Paul trap & \(10^{3}\) &
Particles are launched upwards from a vibrating piezo in the vicinity of a Paul trap. &
Not demonstrated under vacuum conditions. \\ \hline
Khodaee-2022 (Aspelmeyer) \cite{khodaee_dry_2022}& PZT & \shortstack[c]{Silica\\50\,nm} & Dipole trap & \(10^{2}\) &
Particles are put on the lower surface of glass (clamped to piezo) and fall due to piezo vibrations. &
Not demonstrated under vacuum conditions. \\ \hline
Ang-2025 (Jun) \cite{Jerard_Vincent_Rubio_Ang_development}& PZT & \shortstack[c]{Polystyrene\\50\,nm} & Dipole trap & \(10^{3}\) &
Particles are put on the surface of vertical glass (clamped to piezo) and fall due to piezo vibrations. &
Not demonstrated under vacuum conditions. \\ \hline
Delord-2017 (Hétet) \cite{delord_electron_2017} & Electric & \shortstack[c]{Diamond\\10 \(\mu\)m} & Paul trap & \(10^{3}\) &
Particles are put on a copper wire and brought manually in the vicinity of the trap. &
Not demonstrated under vacuum conditions. \\ \hline
Bykov-2019 (Northup) \cite{bykov_direct_2019}& LIAD & \shortstack[c]{Silica\\300\,nm} & Paul trap & \(4\times10^{-7}\) &
Particles are put on a thin metal foil which is irradiated by short laser pulse. The pulse creates acoustic shockwave that ejects the particles. &
High temperatures involved. \\ \hline
Nikkhou-2021 (Millen) \cite{nikkhou_direct_2021} & LIAD & \shortstack[c]{Silica\\300\,nm} & Dipole trap & 1 &
Particles are put on a thin metal foil which is irradiated by short laser pulse. The pulse creates acoustic shockwave that ejects the particles. &
High temperatures involved. \\ \hline
Lindner-2024 (Aspelmeyer) \cite{lindner_hollow-core_2024}& HCF & \shortstack[c]{Silica\\100--755\,nm} & Optical standing-wave trap & \(10^{-9}\) &
Particles are loaded into the HCF via a nebulizer and transported from a low-vacuum chamber into a high-vacuum chamber using two counter propagating beams. In the second chamber they are trapped in a standing-wave trap. &
High intensity laser beams may graphitize and damage NDs. \\ \hline
\end{tabular}
\caption{Summary of nanoparticle loading and trapping experiments with their key limitations.}
\label{tab:loading_trapping}
\end{table}

After exploring the advantages and disadvantages of the above methods to load particles into a Paul trap in UHV conditions, we would like to offer the usage and further exploration of a linear quadrupole electric guide as a new dry source for UHV experiments.
As described in the outlook, a linear guide comprises four parallel electrodes symmetrically placed at a distance $R$ from the center axis . An AC high voltage applied to the electrodes creates two-dimensional confinement while allowing the particles to move freely along the axis parallel to the electrodes. The linear guide has the same design as a linear Paul trap, but without the use of endcaps, which confine the particle in the axial direction. Linear guides are commonly used in mass spectrometry experiments \cite{douglas_linear_2005} (referred to as ion guides in that field). The common use of linear Paul traps in the cold ion community and of linear guides in the mass spectrometry community renders this setup a promising tool for our goal of a clean and efficient particle loading method. Furthermore, this setup was already demonstrated to improve the loading of nanoparticles from the jet of an electrospray apparatus into a Paul trap \cite{alda_trapping_2016, Bullier_optomechanics_2020}. A list of works that use a linear guide to guide nanoparticles is given in Table \ref{tab:electrospray_trapping}. In a work from 2016, \cite{alda_trapping_2016}, the transfer of nanoparticles from a linear guide into a planar Paul trap was demonstrated, after the nanoparticles were loaded into the guide using electrospray. In another relevant work, the trapping of 387\,nm silica particles in a hybrid Paul trap at \SI{e-1}{\milli\bar} was achieved, by guiding the particles from electrospray into the trap \cite{Bullier_optomechanics_2020}.

\begin{table}[htbp]
\centering
\footnotesize
\setlength{\tabcolsep}{3pt}
\renewcommand{\arraystretch}{1.15}
\begin{tabular}{|>{\centering\arraybackslash}m{\cwA}|
                >{\centering\arraybackslash}m{\cwB}|
                >{\centering\arraybackslash}m{\cwC}|
                >{\centering\arraybackslash}m{\cwD}|
                >{\centering\arraybackslash}m{\cwE}|
                >{\centering\arraybackslash}m{\cwF}|
                >{\centering\arraybackslash}m{\cwG}|}
\hline
\shortstack[c]{\textbf{Author--}\\\textbf{Year}\\\textbf{(Group)}} &
\shortstack[c]{\textbf{Loading}\\\textbf{Method}} &
\shortstack[c]{\textbf{Particle's}\\\textbf{Type and}\\\textbf{Size}} &
\shortstack[c]{\textbf{Trapping}\\\textbf{Method}} &
\shortstack[c]{\textbf{Pressure}\\\textbf{[mbar]}} &
\textbf{Details} &
\textbf{Limitations} \\
\hline
Alda-2016 (Quidant) \cite{alda_trapping_2016}&
\shortstack[c]{Electrospray\\into linear\\guide, then\\to trap} &
\shortstack[c]{Polymer\\spheres\\100\,nm} &
Planar Paul trap &
\(10^{3}\) &
The particles are charged with electrospray and guided to the planar Paul trap using a linear Paul guide. &
Wet source (ambient conditions) \\ \hline
Bullier-2020 (Barker) \cite{Bullier_optomechanics_2020}&
\shortstack[c]{Electrospray\\into linear\\guide, then\\to trap} &
\shortstack[c]{Silica\\387\,nm} &
Hybrid Paul trap &
\(10^{-1}\) &
The particles are charged with electrospray at a 2 mbar stage and directed via a beam skimmer into a \(\sim10^{-1}\) mbar chamber, where they are guided via a linear quadrupole guide to a hybrid Paul trap. &
Wet source \\ \hline
\end{tabular}
\caption{Electrospray‐based loading of nanoparticles via linear guides and subsequent trapping, with key operating conditions and limitations. Let us note that some groups load vacuum chambers directly with electrospray. This has several drawbacks such as not being able to go to deep UHV without baking, not being able to go to HV without long-duration pumping, and very long chamber stabilization time (even days) until the chamber is in an electrical steady state.}
\label{tab:electrospray_trapping}
\end{table}

In this work, we suggest a new design for high-efficiency, clean loading of NDs into a Paul trap using a linear guide to transport the nanoparticles from a loading chamber at ambient pressure into a low-pressure science chamber via a differential pumping tube. This approach combines two techniques already proven, the use of a linear guide and the use of a differential pumping tube, where both techniques are commonly used in the fields of levitated nanoparticles, mass spectrometry, and cold atoms. Similarly to the HCF technique that demonstrated direct UHV loading into an optical trap, a linear guide integrated with a concentric differential pumping tube can allow for the transport of a charged nanoparticle from the loading chamber into a UHV science chamber. Combined with the high efficiency of the electric launching in ambient conditions, this source, described in the Outlook, can yield the desired high efficiency for the ND SGI experiments.

\section{Theoretical Background}
\subsection{PZT Method}
A primary challenge in developing a dry source is detaching particles from their initial substrate, which requires overcoming adhesion forces between the particles and the surface. The primary adhesion mechanisms at play include capillary forces, electrostatic double-layer interactions, and van der Waals (VDW) forces. Avoiding moisture suppresses the first two forces. However, VDW forces persist even under UHV conditions and must be actively overcome. The VDW adhesion force between a spherical nanoparticle of radius $r$ and a planar surface with effective surface energy $\gamma$ (the amount of energy per unit area required to separate two materials) can be described using the Derjaguin--Muller--Toporov (DMT) model \cite{derjaguin_effect_1975}:
\begin{equation}
F_{\text{VDW}} = 4\pi\gamma r
\label{eq:Fvdw}
\end{equation}

To detach the particle, the piezo system must generate an inertial force exceeding this adhesion force. The resulting acceleration $a$ required for detachment is derived by dividing the VDW force by the particle's mass $m = \frac{4}{3}\pi r^3 \rho$, where $\rho$ is the mass density of the particle, giving the relation:
\begin{equation}
a = \frac{F_{\text{VDW}}}{m} = \frac{3\gamma}{r^2 \rho}
\label{eq:accel}
\end{equation}
This relation shows that the required acceleration increases as the particle radius decreases. For NDs with density $\rho \approx 3500~\text{kg/m}^3$ and assuming $\gamma \approx 10~\text{mJ/m}^2$ for a diamond-glass substrate (coated with a hydrophobic coating), the necessary acceleration is given by:
\begin{equation}
a \approx \frac{8.57 \times 10^{-6}}{r^2}
\quad \left[\text{m/s}^2\right]
\label{eq:numeric_accel}
\end{equation}

The required acceleration can be achieved by applying high-frequency mechanical vibrations to the surface. Such vibrations are generated by clamping the substrate to a vibrating piezoelectric transducer driven at high frequency. Displacement of the substrate can be modeled as harmonic motion:
\begin{equation}
z(t) = A \sin(\omega t)
\end{equation}
\begin{equation}
a(t) = -A \omega^2 \sin(\omega t)
\label{eq:accel_osc}
\end{equation}
where $A$ is the peak displacement amplitude and $\omega = 2\pi f$ is the angular frequency.

\begin{figure}[h!]
  \centering
  \includegraphics[width=0.8\textwidth]{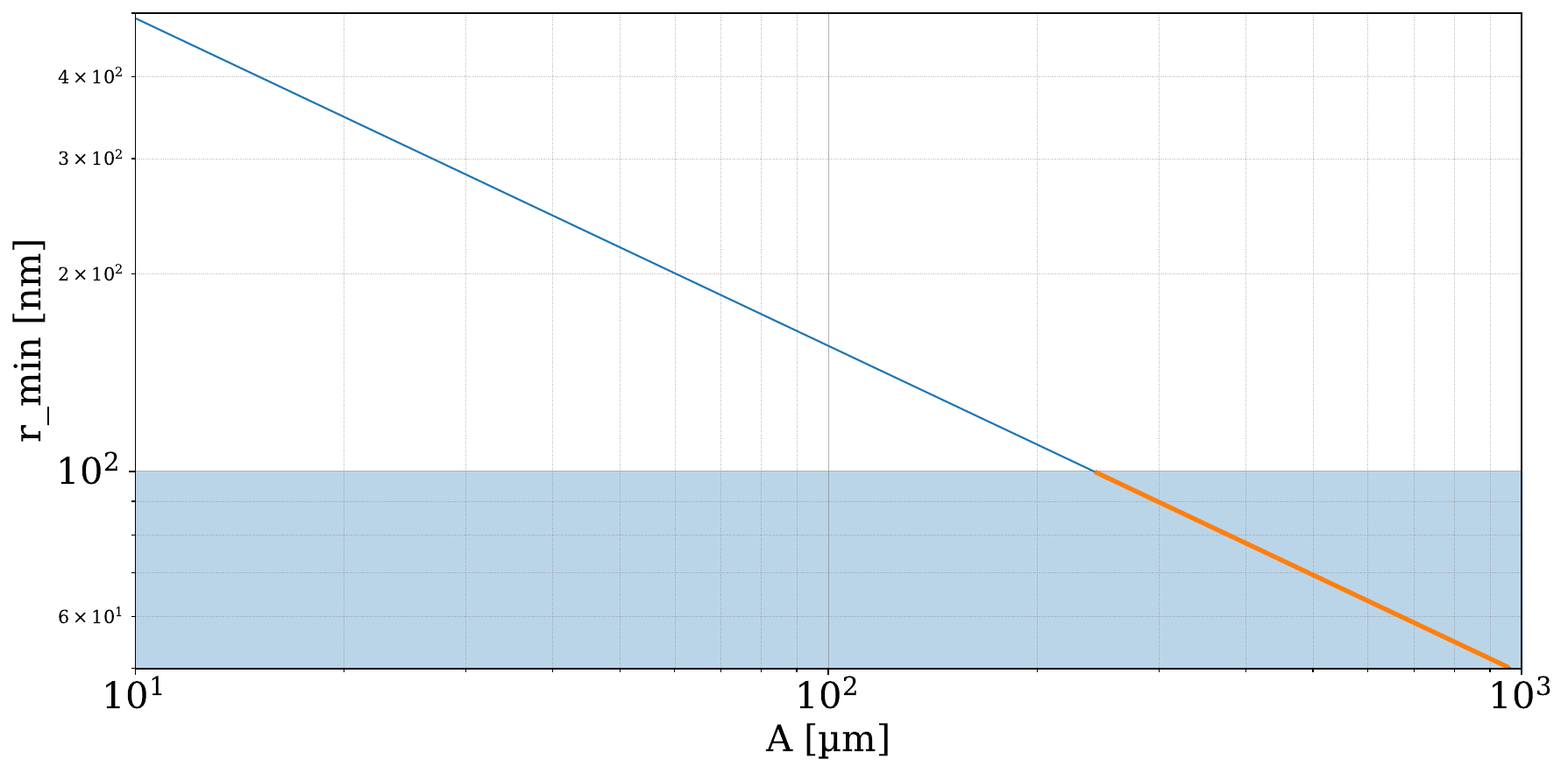} 
  \caption{Graphical representation of Eq.~\eqref{eq:6}, with the given density of diamonds and effective surface energy $\gamma \approx 10~\text{mJ/m}^2$. The red line indicates the minimal particle radius that we are initially interested in for a particle mass of about $10^{-19}$\,kg (about $10^7$ atoms). In our experiments we use a piezo displacement of $A$ whose resonance frequencies can reach 300\,kHz. The figure shows that for example, detaching a nanoparticle with a radius of 50\,nm requires a displacement of nearly 1\,mm of the piezo.}
  \label{fig:Minimal_Launchable_Radius_Of_Particle_At_300_kHz_Resonance_As_Function_Of_Piezo_Displacement}
\end{figure}

Equating the peak acceleration, $A\omega^2$, to the required detachment acceleration gives the minimum launchable particle radius:

\begin{equation}
r_{\text{min}} = \sqrt{ \frac{3\gamma}{A \omega^2 \rho} }
\label{eq:6}
\end{equation}
Eq.~\ref{eq:6} is graphically represented in Fig.~\ref{fig:Minimal_Launchable_Radius_Of_Particle_At_300_kHz_Resonance_As_Function_Of_Piezo_Displacement} showing that the minimal launchable radius decreases with increasing piezo displacement.

\subsection{Electric Launching}
Coulomb's law gives the force exerted by an electric field $\mathbf{E}$ on a charge $q$:

\begin{equation}
\mathbf{F} = q\,\mathbf{E}.
\end{equation}

In quadrupole electric traps, as in a Paul trap, the electric field is an oscillating, spatially inhomogeneous field:
\begin{equation}
\mathbf{E}(\mathbf{r}, t) = \mathbf{E}_0(\mathbf{r}) \cos(\omega t) .
\end{equation}
Assuming the drive frequency $\omega$ is much larger than any secular (slow) frequency of the particle, we may split its trajectory into a slow secular motion $\mathbf{R}(t)$ and a small, fast micromotion $\boldsymbol{\rho}(t)$ \cite{a_v_gaponov_m_a_miller_potential_1957}:
\begin{equation}
\mathbf{r}(t) = \mathbf{R}(t) + \boldsymbol{\rho}(t), \qquad |\boldsymbol{\rho}| \ll |\mathbf{R}| .
\end{equation}
We expand the field about the slow coordinate and retain only the lowest non-trivial order in $\boldsymbol{\rho}$:
\begin{equation}
\mathbf{E}(\mathbf{R} + \boldsymbol{\rho}, t) \simeq \mathbf{E}_0(\mathbf{R}) \cos(\omega t)
+ (\boldsymbol{\rho} \cdot \nabla)\mathbf{E}_0(\mathbf{R}) \cos(\omega t) .
\end{equation}
The equation of motion (neglecting magnetic fields) for a particle with a charge $q$ and mass $m$ is:
\begin{equation}
m \ddot{\mathbf{r}} = q \, \mathbf{E}(\mathbf{r}, t).
\end{equation}

Thus, the micromotion is described by the first order such that:
\begin{equation}
m \ddot{\boldsymbol{\rho}} = q \, \mathbf{E}_0(\mathbf{R}) \cos(\omega t)
\quad \Rightarrow \quad
\boldsymbol{\rho}(t) = - \frac{q}{m \omega^2} \, \mathbf{E}_0(\mathbf{R}) \cos(\omega t) .
\end{equation}
For the secular motion, we average the next-order equation over one RF period $T = 2\pi/\omega$ to remove the first order oscillatory term:
\begin{equation}
m \ddot{\mathbf{R}} = q \Big\langle (\boldsymbol{\rho} \cdot \nabla)\mathbf{E}_0(\mathbf{R}) \cos(\omega t) \Big\rangle
= - \frac{q^2}{2 m \omega^2} \, \nabla \!\left[ E_0^2(\mathbf{R}) \right] .
\end{equation}
Thus, the ponderomotive potential and corresponding force are:
\begin{equation}
\Phi_p(\mathbf{R}) = \frac{q^2}{4 m \omega^2} \, E_0^2(\mathbf{R}),
\qquad
\mathbf{F}_p(\mathbf{R}) = - \nabla \Phi_p(\mathbf{R})
= - \frac{q^2}{4 m \omega^2} \, \nabla E_0^2(\mathbf{R}) .
\end{equation}

In this work, we consider mainly  dielectric non-conductive NDs, which have a low capacity to accumulate electrons. However, under the influence of an electric field, dielectric particles can be polarized, allowing the electric field to exert a force on the dipoles of the neutral particles \cite{pohl_dielectrophoretic_1972}.

We consider a small spherical particle of radius $a$ suspended in a medium.
The applied electric field is spatially inhomogeneous and oscillating with angular frequency $\omega$.

For a homogeneous sphere in a homogeneous medium, the complex polarizability takes the form:
\begin{align}
\alpha(\omega) &= 4\pi\,\varepsilon_m^*(\omega)\,a^{3}\,K(\omega),\\
K(\omega) &= \frac{\varepsilon_p^*(\omega)-\varepsilon_m^*(\omega)}
                 {\varepsilon_p^*(\omega)+2\,\varepsilon_m^*(\omega)},\\
\varepsilon^*(\omega) &= \varepsilon - \frac{i\sigma}{\omega},
\end{align}
where $\varepsilon_p^*$ and $\varepsilon_m^*$ are the complex permittivities of the particle and the medium and $\sigma$ denotes conductivity. The cycle-averaged interaction energy of an induced dipole in a harmonic field is
\begin{equation}
\langle U\rangle = -\frac{1}{4}\,\operatorname{Re}\{\alpha(\omega)\}\,|\tilde{\mathbf{E}}|^{2}.
\end{equation}

Therefore,
\begin{equation}
\langle \mathbf{F}_{\mathrm{DEP}}\rangle = -\nabla \langle U\rangle
= \frac{1}{4}\,\operatorname{Re}\{\alpha(\omega)\}\,\nabla |\tilde{\mathbf{E}}|^{2}.
\end{equation}

The result holds when $a$ is sufficiently small that higher multipoles are negligible and when field variations across the particle are weak.
At very low frequencies in conductive media, $\varepsilon^*$ is dominated by the term $\sigma/\omega$ and $\Re\{K\}$ can change sign with $\omega$, leading to frequency-selective DEP.

Although we are aware of only one work on electric loading of nanoparticles \cite{delord_electron_2017}, the electric launching of microparticles and nanoparticles is widely investigated. A familiar construction for generating a spatially inhomogeneous, oscillating electric field is a configuration of parallel electrodes with a differential voltage applied a differential voltage, thus creating a traveling wave of electric field \cite{zouaghi_assessment_2019,patel_comprehensive_2023}. Such works enabled the investigation of various electric effects on dielectric particles (Coulomb's force, ponderomotive force, dielectrophoretic force).

In our work, we explore several options for applying spatially inhomogeneous, oscillating fields and thus launching particles towards the trap.

\subsection{Paul Trap}
To confine the launched NDs after detachment, we constructed a simple ring Paul trap using a copper ring connected to a high-voltage AC amplifier. The trap operates on the principle of dynamic confinement using an oscillating quadrupole potential, which provides three-dimensional trapping through time-averaged electric forces \cite{paul_electromagnetic_1990}.

For a Paul trap with an ideal quadrupole potential including both a DC voltage \(U\) and a RF voltage \(V\cos(\Omega t)\) the potential is:

\begin{equation}
\Phi(x,y,z,t)=\frac{U+V\cos(\Omega t)}{2\,R^{2}}\!\bigl(x^{2}+y^{2}-2z^{2}\bigr).
\end{equation}

For a particle of mass \(m\) and charge \(Q\) the Newton equation \(m\ddot{\mathbf r}+Q\nabla\Phi=0\) becomes, after setting \(\tau=\Omega t/2\):

\begin{equation}
\frac{d^{2}x_{i}}{d\tau^{2}}+\bigl(a_{i}-2q_{i}\cos(2\tau)\bigr)x_{i}=0,
\qquad (i=x,y,z),
\end{equation}

where the dimensionless Mathieu parameters are:  

\begin{equation}
\displaystyle
a_x = a_y = -\frac{a_z}{2} = -\frac{4QU}{m\Omega^{2}R^{2}},
\qquad
q_x = q_y = -\frac{q_z}{2} = \frac{2QV}{m\Omega^{2}R^{2}}
\end{equation}

These six Mathieu parameters define the stability regions of the trap. In this experiment, the DC bias is set to zero $U = 0$, yielding: \(a_{i}=0\). Thus, the appropriate stability region for the $q$ parameters are:

\begin{equation}
|q|\;\le\;0.908
\quad\Longleftrightarrow\quad
\left|\,\frac{2QV}{m\Omega^{2}R^{2}}\,\right|\le0.908.
\end{equation}

\subsection{Linear Guide}
\label{subsection:Linear_Guide}

For four parallel electrodes on a circle of radius \(R\), where opposite electrodes are driven with the same RF phase, and adjacent electrodes with opposite RF phase, near the trap axis, including both a DC voltage \(U\) and RF voltage \(V\cos(\Omega t)\), the potential can be approximated by a quadrupole potential \cite{paul_electromagnetic_1990}:

\begin{equation}
\Phi(x,y,t)=\frac{U+V\cos(\Omega t)}{2\,R^{2}}\bigl(x^{2}-y^{2}\bigr),
\end{equation}
For a particle of mass \(m\) and charge \(Q\) the Newton equation \(m\ddot{\mathbf r}+Q\nabla\Phi=0\) becomes, after setting \(\tau=\Omega t/2\):

\begin{equation}
\frac{d^{2}x_{i}}{d\tau^{2}}+\bigl(a_{i}-2q_{i}\cos(2\tau)\bigr)x_{i}=0,
\qquad (i=x,y),
\end{equation}

where the dimensionless Mathieu parameters are:    

\begin{equation}
\displaystyle
a_x = -a_y = \frac{4QU}{m\Omega^{2}R^{2}},
\qquad
q_x = -q_y = \frac{2 QV}{m\Omega^{2}R^{2}}
\end{equation}

These four Mathieu parameters define the stability regions of the trap. In this experiment, the DC bias is  set to zero $U = 0$, yielding: \(a_{i}=0\). Thus the appropriate stability region for the $q$ parameters is:

\begin{equation}
\label{eq:stability_parameter}
|q|\;\le\;0.908
\quad\Longleftrightarrow\quad
\left|\,\frac{2QV}{m\Omega^{2}R^{2}}\,\right|\le0.908.
\end{equation}

The lowest-order solutions for the motion equation, 
modulated by intrinsic micromotion, are:
\begin{align}
x(t) &= x_{0}\cos\!\bigl(\omega_{x} t+\phi_{x}\bigr)\left(1+\frac{q_{x}}{2}\cos\Omega t\right),\\
y(t) &= y_{0}\cos\!\bigl(\omega_{y} t+\phi_{y}\bigr)\left(1+\frac{q_{y}}{2}\cos\Omega t\right),
\end{align}
where the phases $\phi_{x}$ and $\phi_{y}$ can be set to zero and $\omega_x, \omega_y$ are the secular frequencies:

\begin{equation}
\omega_{x}=\frac{\Omega}{2}\sqrt{a_{x}+\frac{q_{x}^{2}}{2}},
\qquad
\omega_{y}=\frac{\Omega}{2}\sqrt{a_{y}+\frac{q_{y}^{2}}{2}}.
\end{equation}

In the adiabatic approximation we can approximate the secular motion as harmonic, thus using the equipartition theorem:

\begin{equation}
k_{B}T=\frac{1}{2}m\omega_{x}^{2}\langle x_{0} \rangle^{2}
+\frac{1}{2}m\omega_{y}^{2}\langle y_{0} \rangle^{2}.
\end{equation}

Where $k_B$ is the Boltzmann constant, $T$ is temperature and $m$ particle's mass.
Since $a_{x}=a_{y}=0$ and ${q_{x}^{2}}={q_{y}^{2}}\equiv {q}^{2}$, so
$\omega_{x}=\omega_{y}\equiv\omega$.
Defining the radial secular RMS:
$\langle r_{0} \rangle = \sqrt{\langle x_{0} \rangle^{2} + \langle y_{0} \rangle^{2}}$, we can write:

\begin{equation}\label{eq:secular_motion_amplitude}
k_{B}T=\frac{1}{2}m\omega^{2}\langle r_{0} \rangle^{2}
\quad\Rightarrow\quad
\langle r_{0} \rangle=\sqrt{\frac{2k_{B}T}{m\omega^{2}}}
\quad\Rightarrow\quad
r_{0}=\sqrt{\frac{4k_{B}T}{m\omega^{2}}}
\end{equation}

\section{Experimental Setup}

\subsection{Piezoelectric Ring and Slide}
\begin{figure}[h!]
  \centering
  \includegraphics[width=0.8\textwidth, trim=0cm 2cm 0cm 0cm, clip]{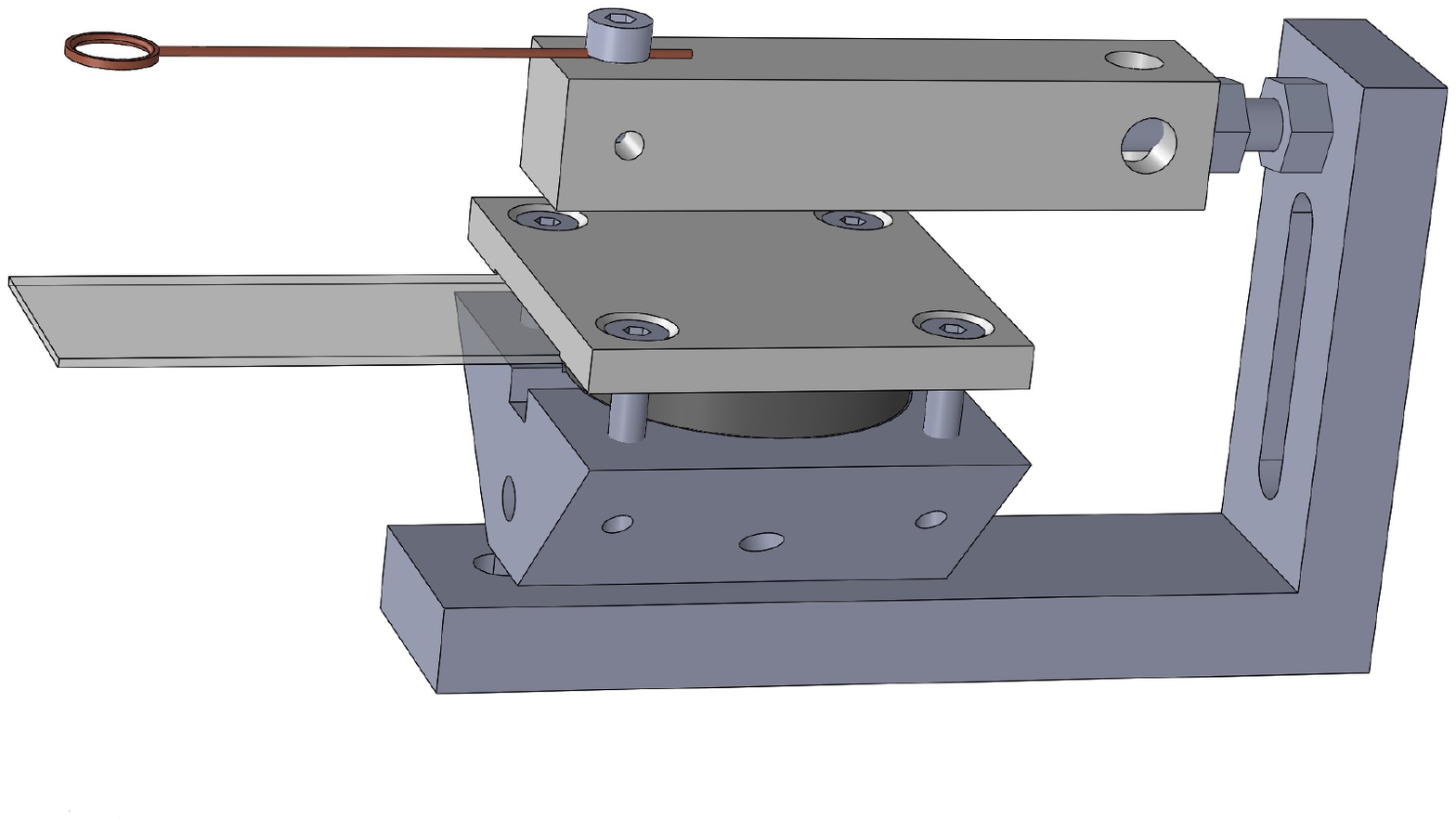}
  \caption{3D model of the piezo ring clamped to a glass slide holding the particles and the Paul-trap ring above. The Paul-trap ring is put above the slide so that the particles arrive with zero velocity, namely, no further dissipation of energy is required.}
  \label{fig:piezo_stage}
\end{figure}

The mechanical setup is based on a ring-shaped piezoelectric transducer (APC International, Material 840), with an outer diameter 38~mm, inner diameter 13.2~mm, and thickness 6.4~mm. The piezo was axially soldered to two electrodes and mechanically supported from below by an aluminum base. A standard microscope ITO-coated (PGO, No. CEC500S) glass slide (76~mm $\times$ 27~mm $\times$ 1--2~mm) is clamped tightly on top of the piezo by a PEEK base, as shown in Fig.~\ref{fig:piezo_stage}. In addition, we spin-coated the glass with a hydrophobic conductive coating to reduce adhesive forces. The cover was "FluoroPel 600", 0.2\% concentration (Cytonix) and it was applied to the glass by applied by spin coating for 30 seconds at a speed of 1000\,rpm. This product is suitable because it keeps the slide at a surface energy of $8\,\mathrm{mJ}\,\mathrm{m}^{-2}$
  and intrinsically conductive for cover thickness smaller than 500\,nm, thereby preserving the particle charging mechanism.

Continuous operation at high voltage and frequency (e.g., 100--300\,V at 300--350\,kHz) led to significant heating of the transistor and surrounding components. To mitigate this, the system was configured to operate in burst mode: voltage was applied in short pulses of 50--100\,ms duration, followed by a cooling interval of approximately 1\,s. This duty cycle significantly reduced the thermal load while preserving resonant excitation efficiency.
To drive the piezoelectric transducer at high frequency and voltage, we constructed a custom electronic circuit design. The core of the circuit consists of a high-voltage transistor acting as a switch, controlled by a function generator outputting a square wave. A series capacitor was used to block DC bias and allow efficient AC transmission.
The power supply used in the experiments was capable of providing up to 300\,V and 1\,A. In continuous mode, the circuit reached the 1\,A current limit at only 100~V when operating at 350\,kHz, which limited the performance. However, in burst mode, we were able to achieve the full 300\,V output even at high frequencies, without exceeding the current threshold.

\begin{figure}[h!]
  \centering
  \includegraphics[width=0.8\textwidth]{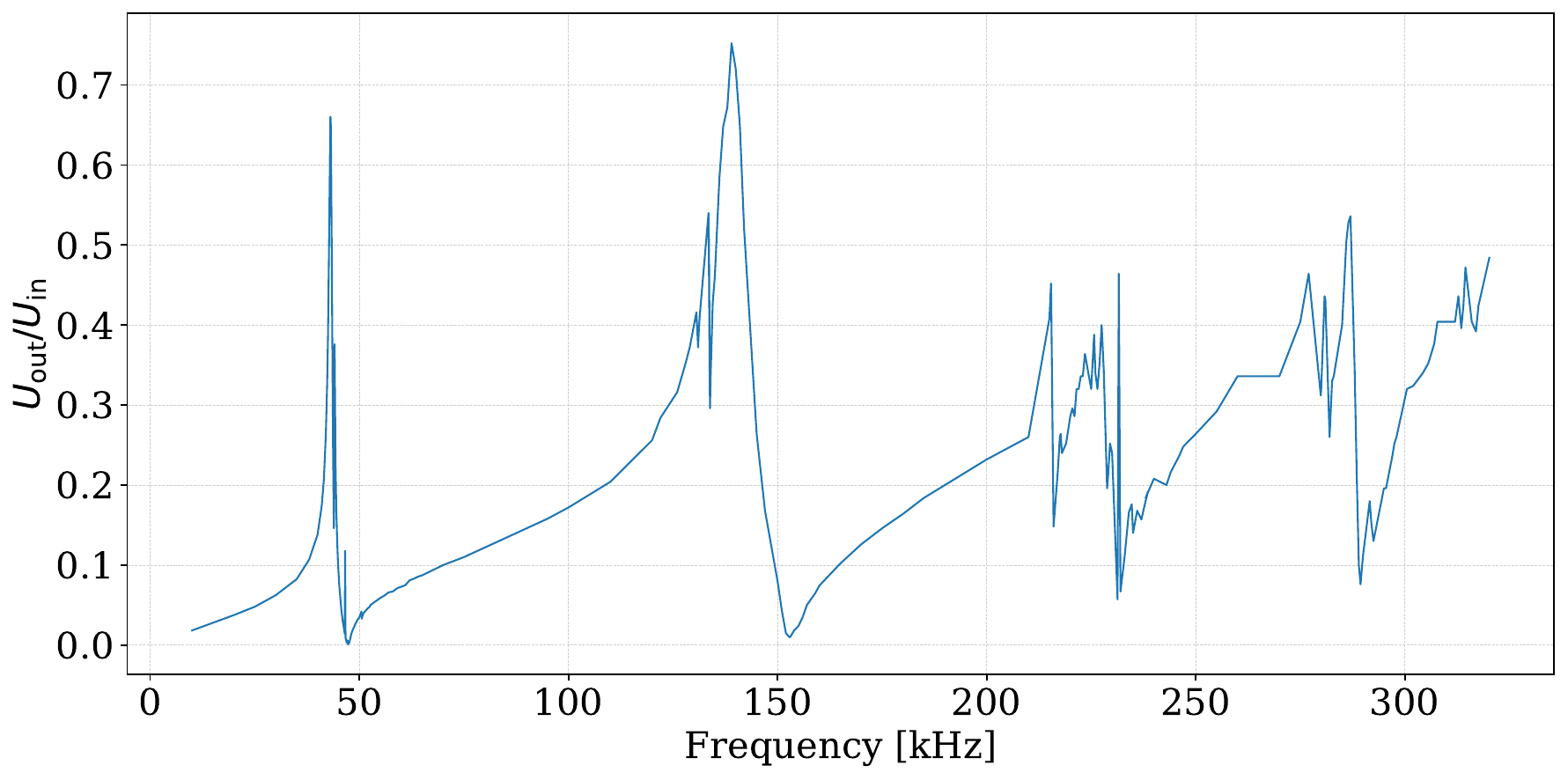}
 \caption{Normalized output-to-input voltage ratio \(U_{\text{out}}/U_{\text{in}}\) of the piezoelectric transducer as a function of excitation frequency. Prominent peaks correspond to mechanical resonance modes of the device, where electromechanical energy conversion is maximized. Minima represent anti-resonances, at which the mechanical response suppresses energy transfer to the measurement resistor.}
\label{fig:resonances}
\end{figure}

To investigate the mechanical resonances of the piezoelectric element, we performed a frequency sweep in which the transducer was driven by an AC voltage while its electrical response across a measurement resistor was recorded. Due to the electromechanical coupling of piezoelectric materials, mechanical vibration modes are reflected as characteristic features in the electrical transfer function. Resonances occur when the driving frequency matches a natural vibrational mode of the structure, leading to enhanced energy conversion from electrical to mechanical form and a corresponding increase in the measured output signal. By normalizing \(U_{\text{out}}\) to the applied \(U_{\text{in}}\) at each frequency, the resulting spectrum (Fig.~\ref{fig:resonances}) isolates the frequency-dependent transfer efficiency of the device, enabling the identification of fundamental and higher-order vibrational modes.

Precise characterization of the piezo-induced displacement is essential for estimating the mechanical acceleration experienced by particles during launch. We developed an optical method to measure the vibration amplitude of the glass slide with high sensitivity, as shown in Fig.\,\ref{fig:optical_detection}:
\begin{figure}[h!]
  \centering
  \includegraphics[width=0.8\textwidth]{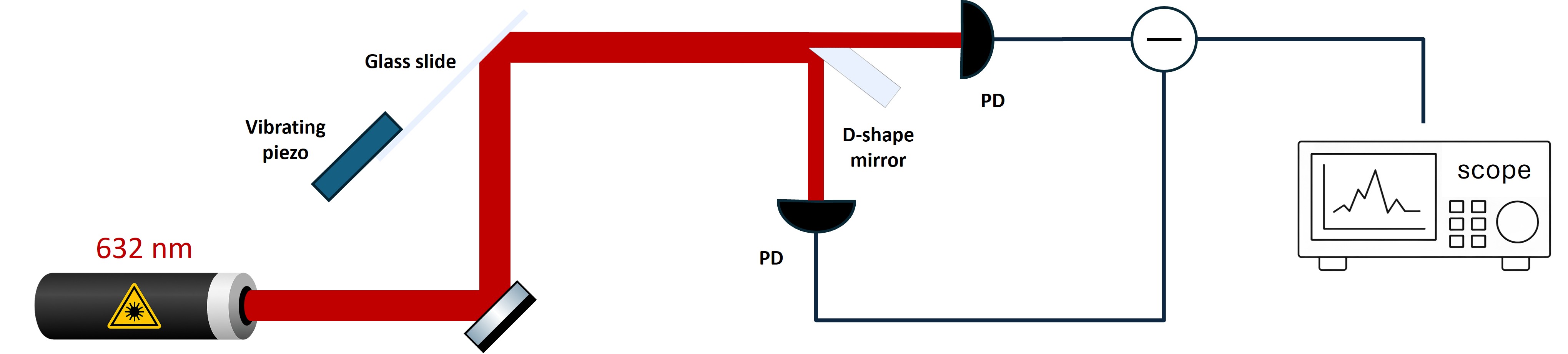}
  \caption{Optical detection setup for measuring the vibration amplitude of the glass slide driven by the piezo. A 632\,nm red laser is directed onto the vibrating slide, and the reflected beam is split by a D-shaped mirror between two photodetectors (PDs). The differential signal from the PDs, processed by a differential amplifier, provides a sensitive measure of slide displacement. This configuration suppresses common-mode noise and enables precise, real-time monitoring of vibration amplitude as a function of drive frequency and voltage.}
  \label{fig:optical_detection}
\end{figure}

To calibrate the system, the entire piezo-glass assembly was mounted on a precision translation stage. Signal vs. position was measured at several points within the range of known displacements of 100~\textmu m to correlate the differential signal to physical motion. This setup enabled real-time measurement of vibration amplitude as a function of drive frequency and voltage.

\begin{figure}[h!]
  \centering
  \includegraphics[width=0.8\textwidth]{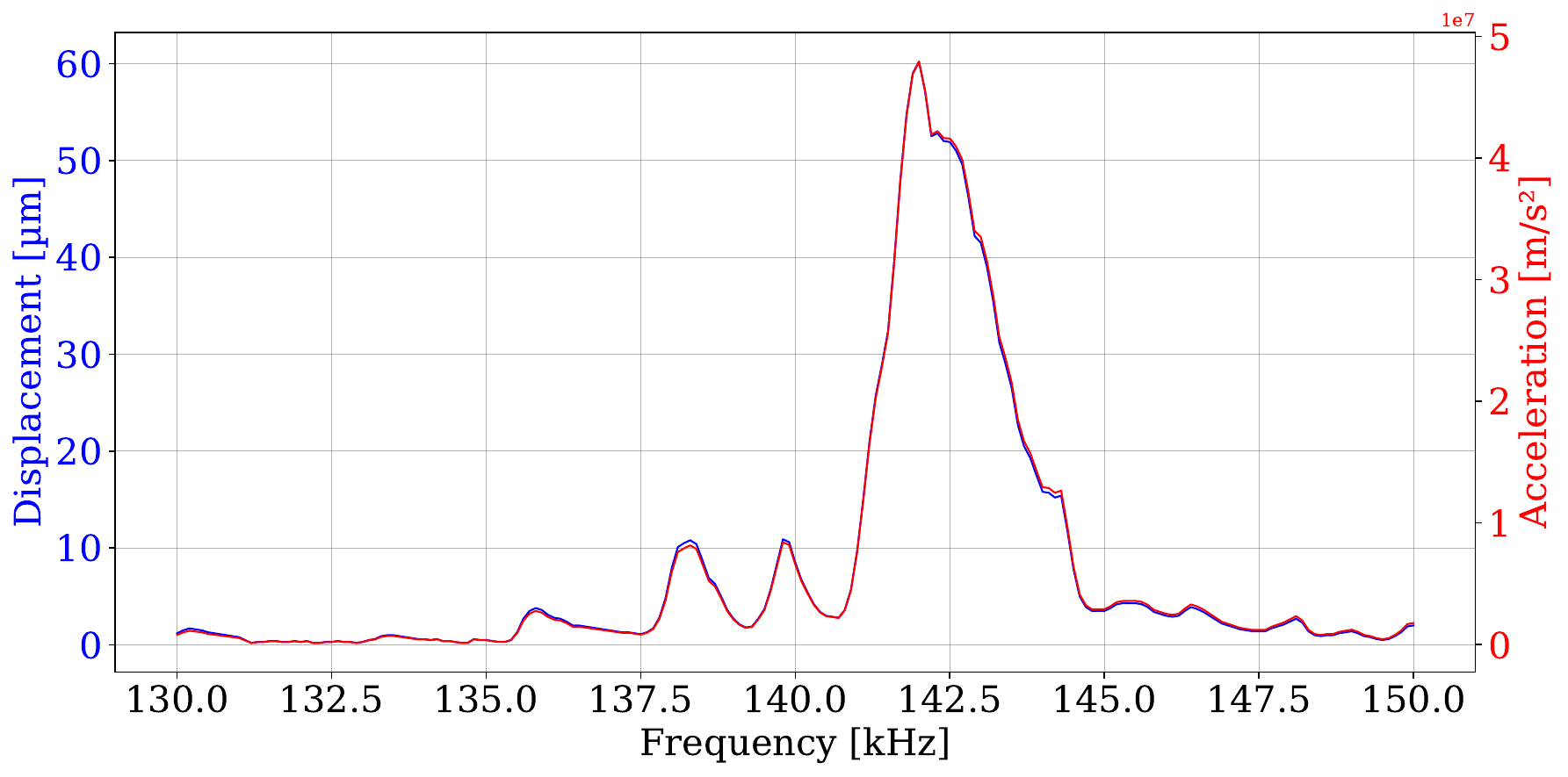}
  \caption{Frequency response of the vibrating slide around the second mechanical resonance mode at approximately 142\,kHz. The plot shows both the measured displacement amplitude and the corresponding peak acceleration as the drive frequency is swept. A sharp peak in both quantities indicates strong mechanical amplification at resonance.}
  \label{fig:displacement_and_acceleration_as_function_of_frequency_around_second_resonance}
\end{figure}

\begin{figure}[h!]
  \centering
  \includegraphics[width=0.8\textwidth]{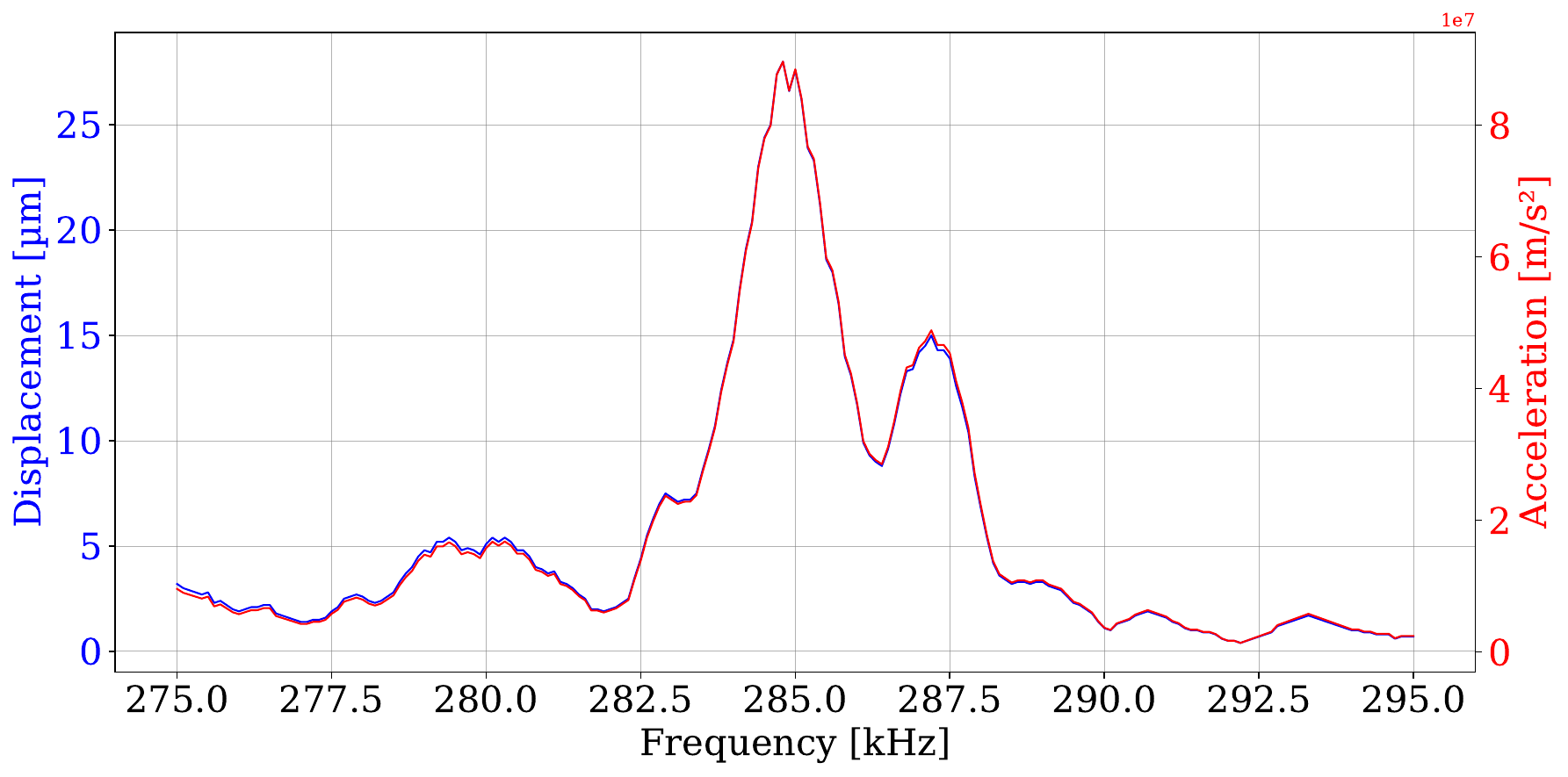}
  \caption{Frequency response around the third mechanical resonance mode of the slide, centered at approximately 285\,kHz. Compared to the first mode, the maximum displacement is roughly half, while the corresponding acceleration is higher due to the increased frequency. A sharp peak in both quantities indicates strong mechanical amplification at resonance.}
  \label{fig:displacement_and_acceleration_as_function_of_frequency_around_third_resonance}
\end{figure}

Since acceleration exhibits a quadratic dependence on frequency, we scanned the drive frequency in the high frequency range of 100 to 350\,kHz and identified several resonance peaks corresponding to strong mechanical amplification. Using a fixed peak voltage of 30\,V, we measured displacement at two key resonance frequencies: 142\,kHz, as shown in Fig.~\ref{fig:displacement_and_acceleration_as_function_of_frequency_around_second_resonance}, and 285\,kHz, as shown in Fig.~\ref{fig:displacement_and_acceleration_as_function_of_frequency_around_third_resonance}.

\begin{figure}[h!]
  \centering
  \includegraphics[width=0.8\textwidth]{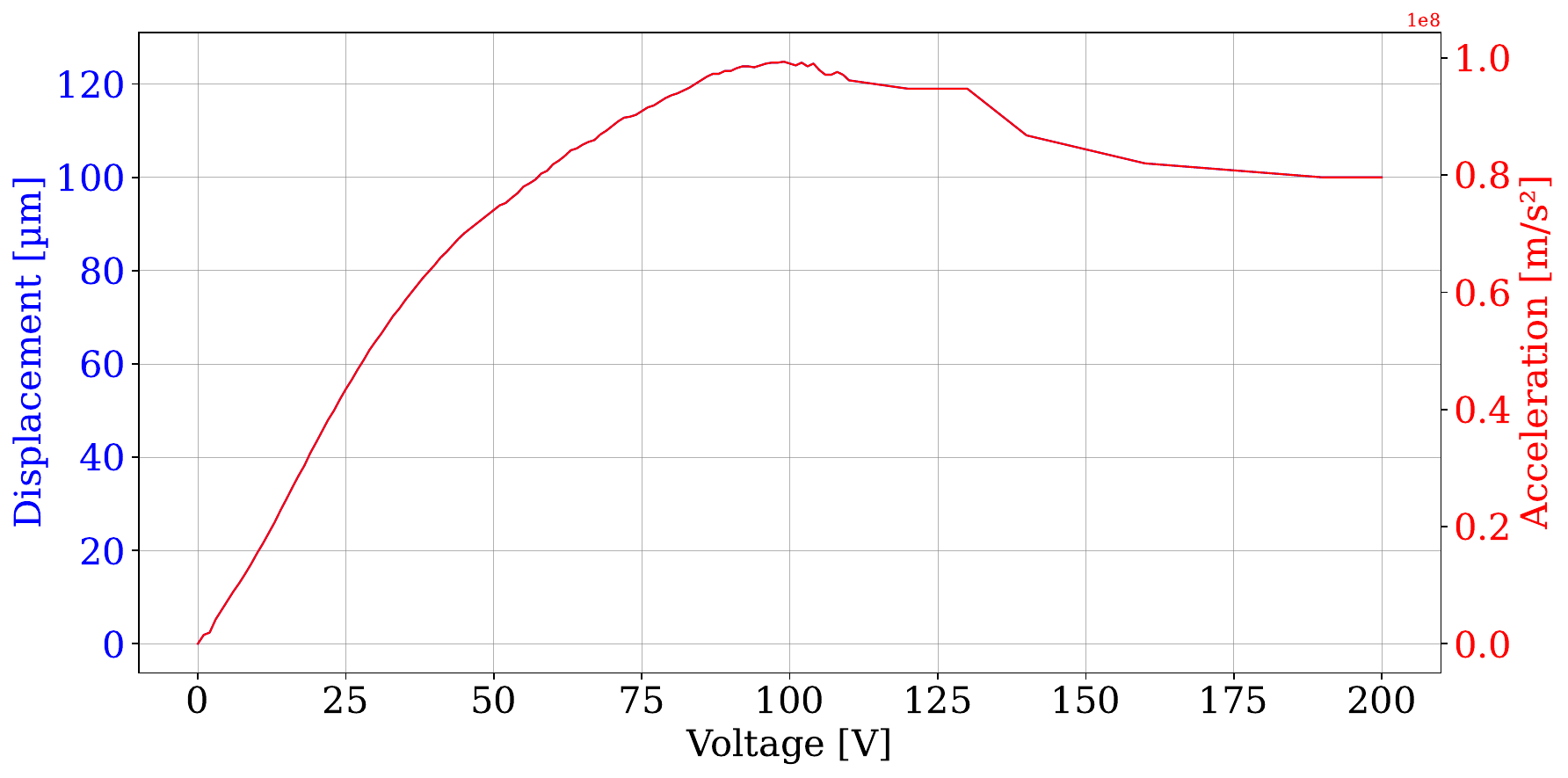}
  \caption{Dependence of displacement and acceleration on the applied drive voltage for the second resonance mode at 142\,kHz, using short bursts of 10{,}000 cycles with 1\,s rest intervals to avoid heating effects. The displacement increases nearly linearly at low voltages before saturating at around 90\,V, with the acceleration following the same trend.}
  \label{fig:Displacement_and_acceleretion_as_function_of_voltage_at_second_resonance}
\end{figure}

\begin{figure}[h!]
  \centering
\includegraphics[width=0.8\textwidth]{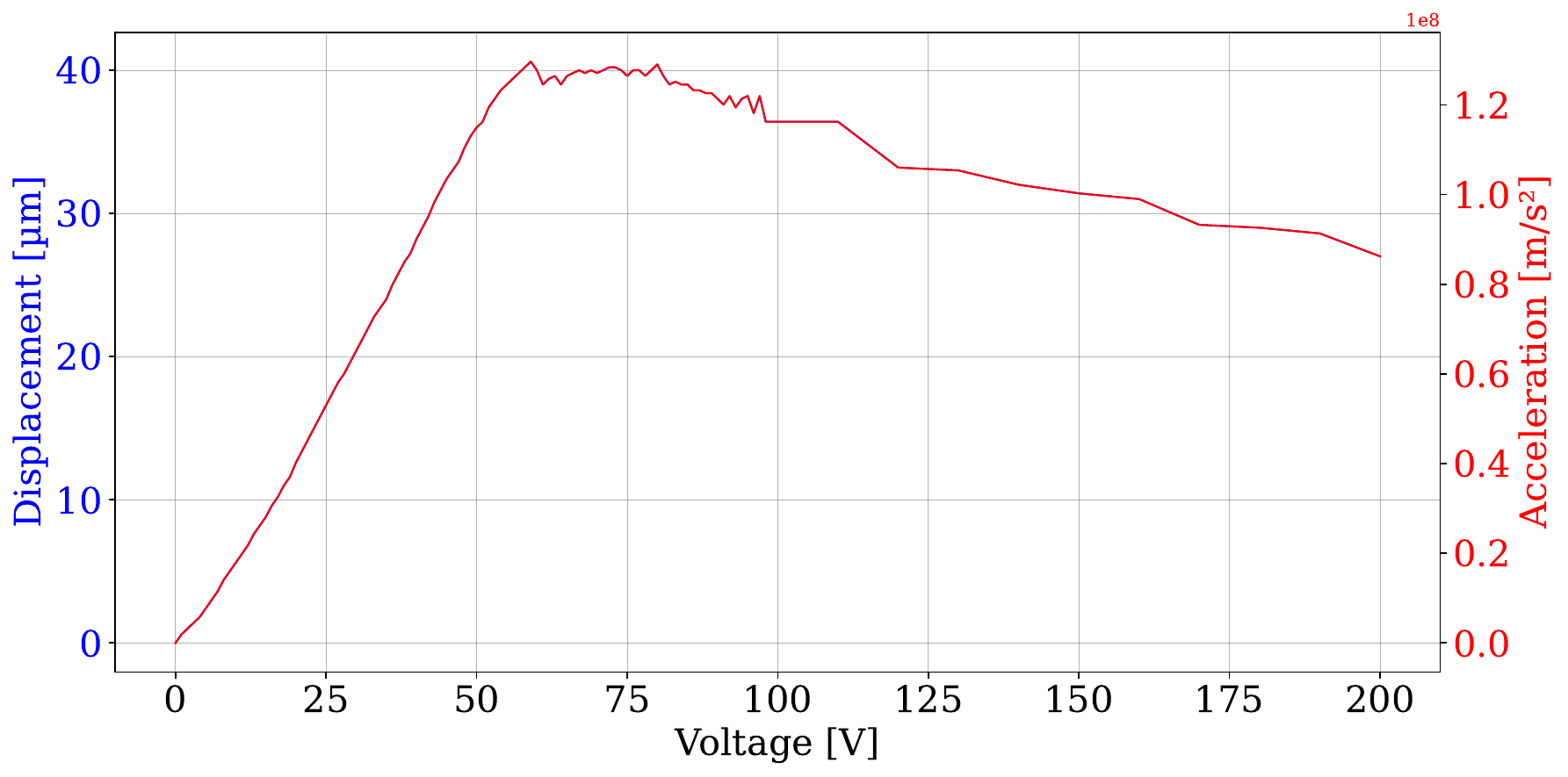}
 \caption{Voltage dependence of displacement and acceleration at the third resonance mode (285\,kHz), measured with the same pulsed excitation scheme as in Fig.~\ref{fig:Displacement_and_acceleretion_as_function_of_voltage_at_second_resonance}. Saturation occurs at a lower voltage of about 60\,V, indicating that higher-frequency modes reach their maximum amplitude at reduced drive levels, which in turn limits the maximum achievable acceleration compared to the first mode.}
 \label{fig:displacement_and_acceleration_as_function_of_voltage_at_third_resonance}
\end{figure}

As expected, the displacement decreased with increasing frequency. We then investigated the voltage dependence of displacement at each resonance, using short bursts of excitation to avoid thermal saturation. Pulses of 10,000 cycles were applied with a rest period of 1\,s between each burst. The results for the 142\,kHz resonance are presented in Fig.~\ref{fig:Displacement_and_acceleretion_as_function_of_voltage_at_second_resonance}, and those for the 285\,kHz resonance are shown in Fig.~\ref{fig:displacement_and_acceleration_as_function_of_voltage_at_third_resonance}.

Despite the capability of the power supply to deliver voltage exceeding 100\,V, we observe a saturation in displacement at 90\,V for the second resonance and at 60\,V for the third resonance. Furthermore, at higher-frequency modes the saturation occurred at a lower voltage, thus limiting the maximum achievable acceleration.

We also studied the effect of pulse duration on saturation. At a fixed voltage of 90\,V and resonance of 142\,kHz, we varied the number of cycles per pulse. Displacement saturated for pulse lengths shorter than approximately 21\,ms (about 3,000 cycles), indicating a minimum duration required for maximum amplitude.

Despite the capability of the power supply to deliver voltages exceeding 100\,V, the observed saturation thresholds---90\,V for the second resonance and 60\,V for the third---confirm this trend. Furthermore, for higher-frequency modes, saturation occurs at a lower voltage, further constraining the maximum achievable acceleration.

We also studied the effect of pulse duration on saturation. At a fixed voltage of 90\,V and a resonance frequency of 142\,kHz, we varied the number of cycles per pulse. The results showed that displacement saturates for pulse lengths shorter than approximately 21\,ms (about 3{,}000 cycles), indicating the minimum duration required to reach maximum amplitude.

\subsection{Particles}

\begin{figure}[htbp]
  \centering
  \begin{subfigure}[t]{0.48\linewidth}
    \centering
    \includegraphics[width=\linewidth]{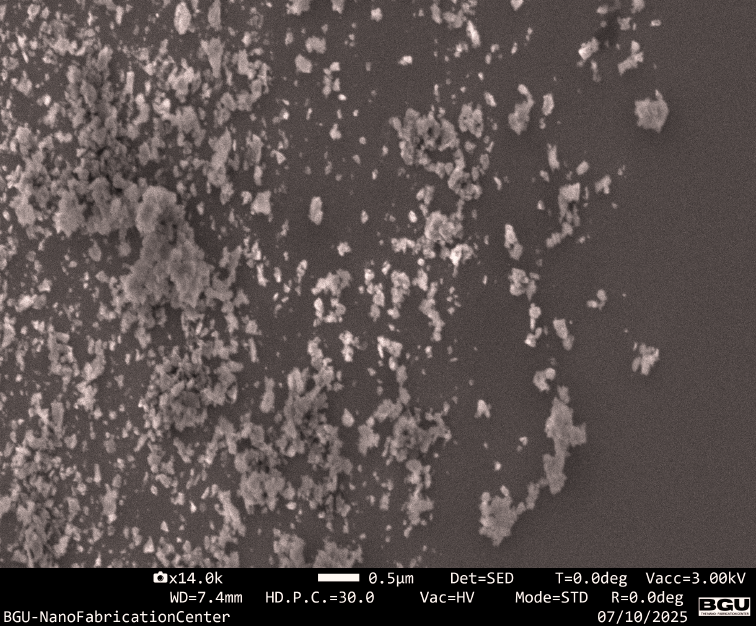}
  \end{subfigure}\hfill
  \begin{subfigure}[t]{0.48\linewidth}
    \centering
    \includegraphics[width=\linewidth]{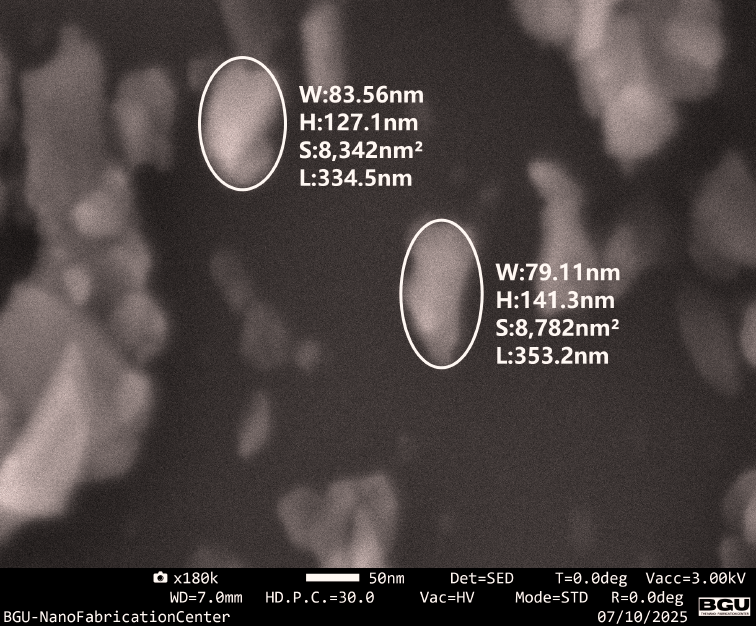}
  \end{subfigure}
  \caption{SEM images of monocrystalline ND powder from Pureon (product No.\,MSY 0-0.15, nominal median diameter 75\,nm) deposited on a conductive substrate. The left panel shows a representative sample illustrating the wide particle size distribution and clustering tendency. The right panel shows measurements of individual particle dimensions from their 2D projections: $L$ denotes the ellipse major axis length, $W$ the ellipse minor axis length, $S$ denotes the projected area of the 3D particle and $H$ denotes its  perimeter. The powder was baked at \SI{100}{\celsius} for three hours before imaging to reduce moisture-induced clustering.}
  \label{fig:SEM_pictures}
\end{figure}

In the experiment we used monocrystalline ND powder supplied by Pureon (product No.\,MSY 0-0.15) with a nominal median diameter of 75\,nm. The particles were deposited onto a conductive substrate and examined using scanning electron microscopy (SEM), as shown in Fig.~\ref{fig:SEM_pictures}. One challenge when working with a dry particle source is preventing clustering, which is mainly caused by moisture. To minimize this effect, the powder was baked at \SI{100}{\celsius} for three hours prior to measurements. In the size analysis presented in the right panel of Fig.~\ref{fig:SEM_pictures}, the parameter $L$ denotes the major axis length of the particle’s projected 2D image, while $H$ denotes the minor axis length, corresponding to the shorter dimension in the image plane.

\section{Results}

\label{section:Results}
We place a sample of 75\,nm NDs on the edge of the glass slide and a copper ring (9\,mm inner diameter, 11\,mm outer diameter) 10\,mm above the slide with 45 degrees between the normal of the ring plane and the normal to the slide, thus enabling effective imaging and optical control. We use a simple CCD camera for imaging and a 532\,nm green laser for scattering. The ring is connected to high voltage amplifier and driven with an RF signal of 5\,kV at 200\,Hz. The glass slide is clamped to a vibrating piezo which is driven at a resonance frequency of 142\,kHz with 90\,V in burst mode, with 1\,s gap between every 10,000 cycles (a 70\,ms duty cycle). 

\begin{figure}[htbp]
  \centering
  \begin{subfigure}[t]{0.32\linewidth}
    \centering
    \includegraphics[width=\linewidth]{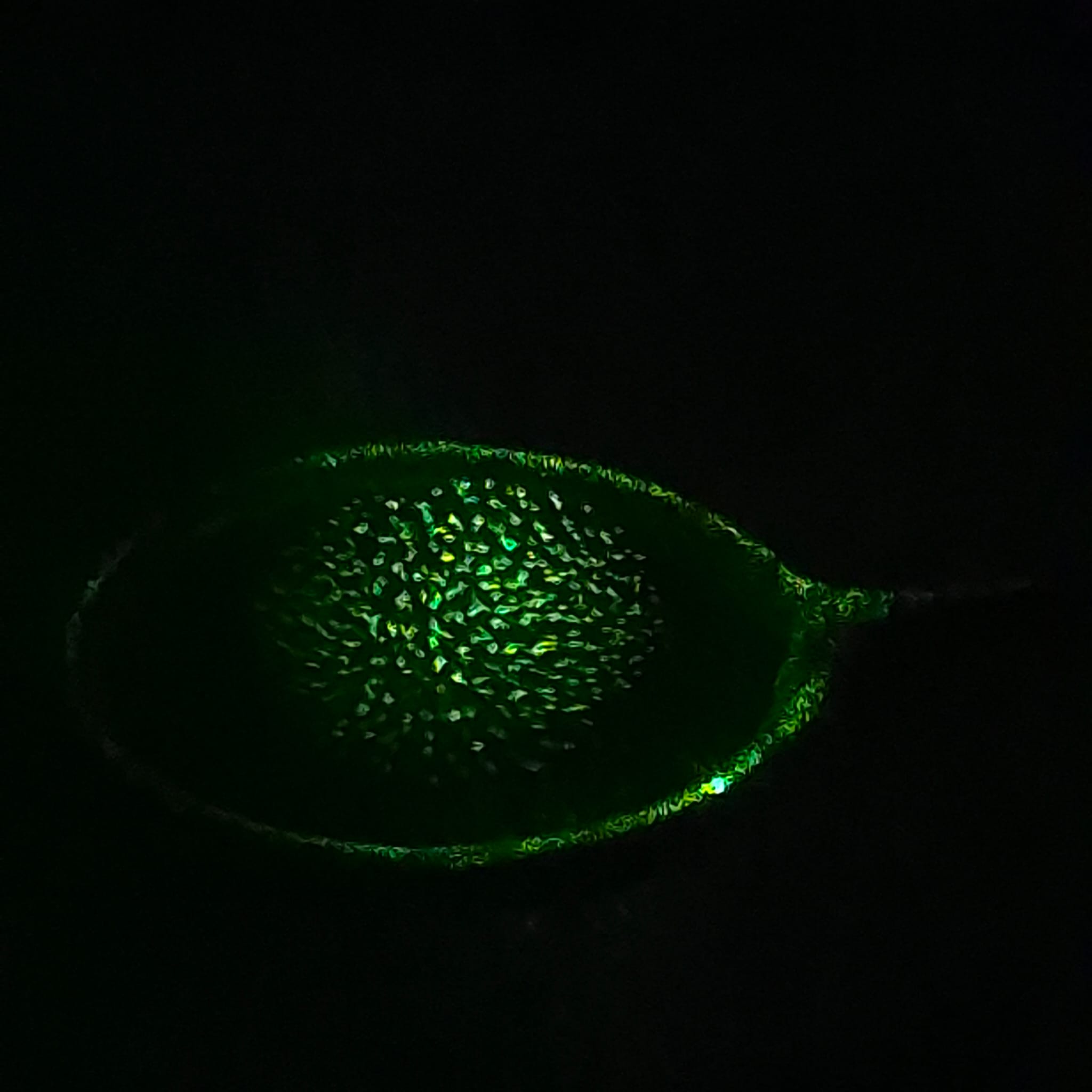}
  \end{subfigure}\hfill
  \begin{subfigure}[t]{0.32\linewidth}
    \centering
    \includegraphics[width=\linewidth]{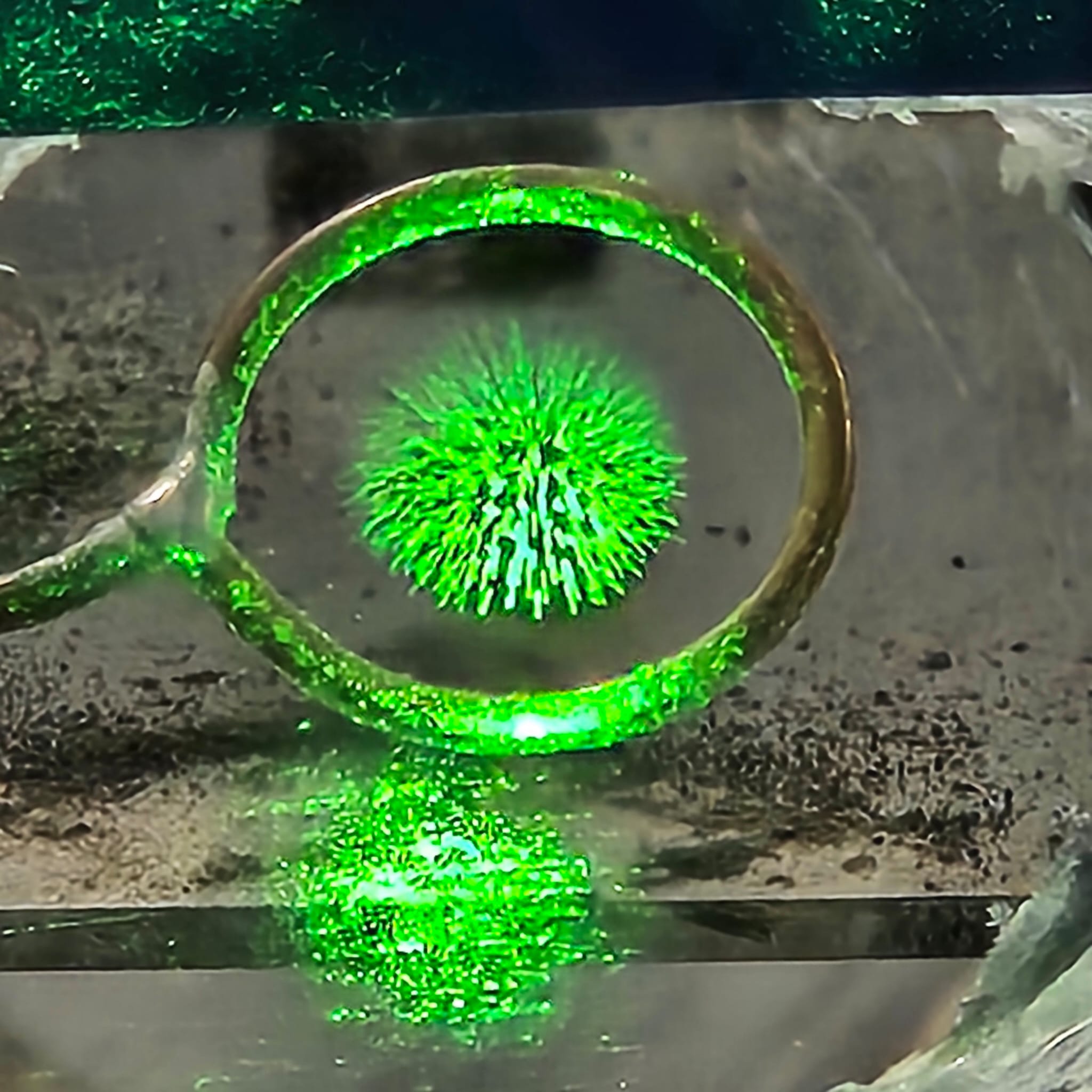}
  \end{subfigure}\hfill
  \begin{subfigure}[t]{0.32\linewidth}
    \centering
    \includegraphics[width=\linewidth]{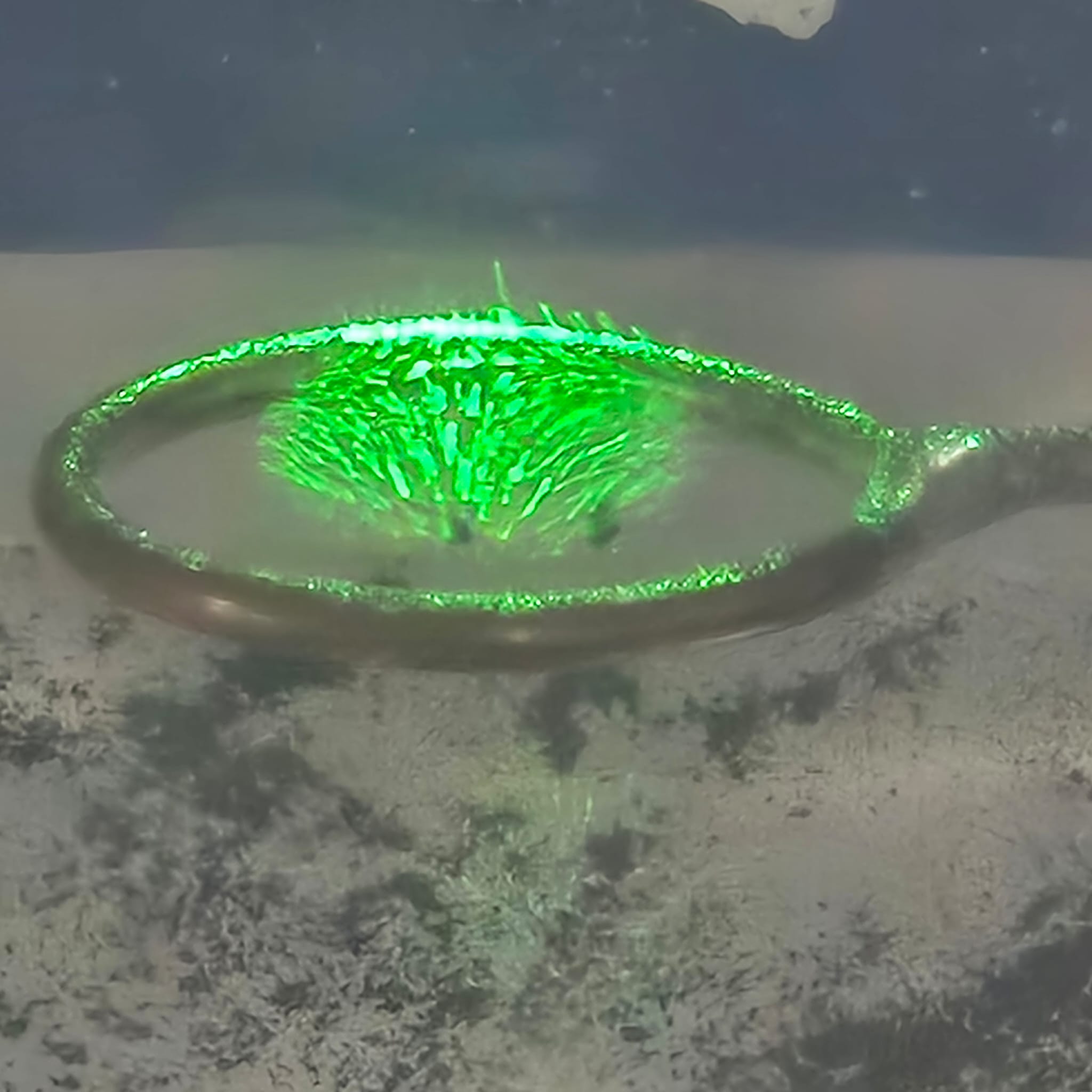}
  \end{subfigure}
  \caption{Trapping of 75\,nm NDs in a Paul trap in ambient conditions. A ring Paul trap with a diameter of 9\,mm is placed approximately 10\,mm at a $45^{\circ}$ angle above a grounded conducting surface, in this case a glass slide coated with ITO. Once the high-voltage AC signal is applied to the ring, we observe a flow of particles from the surface into the trap. The loading effect starts at a voltage of approximately 5\,kV and driving frequency of 140\,Hz, while the trapping itself remains stable at a lower voltage 2.5\,kV. We have a good reason to assume that the electric breakdown of air has a role in the launching of the particles, as both audible and visual signs of breakdown were correlated with the loading of particles into the trap. In a vacuum of $10^{-5}\,\rm mbar$, such a setup did not show any loading, again suggesting that electric breakdown of air has a significant role in the launching and loading.}\label{fig:three}
\end{figure}

Although our goal is to develop a UHV loading method, we first test our system under ambient conditions. We report the successful trapping of tens of particles in each attempt in ambient conditions, as shown in Fig~\ref{fig:three}. Moreover, turning off the piezo and turning on the ring voltage, produces the same large-scale and efficient loading pattern, leading us to the conclusion that the ring itself is launching and trapping the particles. This is a clear demonstration of electric launching and loading by a simple mechanism that requires only the Paul itself and a nearby conducting surface on which the particles are placed. We also observed a similar effect when using a linear Paul trap, with a conducting surface placed below the trap. On top of that, we have preliminary results that show that the efficiency of this loading mechanism is very high, on the order of 1-10\%. We test the efficiency by trapping a few tens of particles in the Paul trap and releasing them above a clean conducting surface. By ensuring that the surface is clean and contains only the particles we released, we can estimate the efficiency. After the release, we turn on the trap again and observe a few particles being trapped, thus we can estimate an efficiency of 1-10\%.

However, carrying out the same experiment (ring voltage on, piezo off) inside the vacuum chamber at a pressure of \SI{e-5}{\milli\bar} demonstrates no launching and no trapping, indicating that the electrical phenomena observed in air is probably a result of electric breakdown of the air. We assume that the high voltage ring, placed close to a conducting slide, creates an image charge below the slide, thus inducing a high electric field, electric breakdown of the air, air ionization accompanied by particle charging and electric attraction to the ring, resulting in high efficiency trapping in conservative trap aided by air damping. Moreover, activating the piezo while turning off the ring voltage yields only vibrations of the particles, indicating that the piezo vibrations alone can't launch particles, but only to detach them from the surface.

Following these conclusions, we face three main challenges in vacuum conditions: charging, launching and trapping. In a recent study about electric forces on dielectric particles, particle charging was demonstrated by placing the particles on a vibrating feeder, which increased friction between the particles and the feeder, thus enhancing the triboelectric effect \cite{zouaghi_assessment_2019}. Thus, we conclude that the piezo is still suitable in vacuum conditions for vibrating the particles and charging them. Indeed, while activating the piezo vibrations and applying high voltage to the ring, we observe massive launching of particles, confirming the success of the charging method. However, the exact amount of charge on the NDs is not measured, and as was found in \cite{fonseca_nonlinear_2016}, it may be limited to just a few elementary charges, which would make it hard to trap the NDs. While operating the piezo and applying high voltage to the ring, we observe a high-flux launching of particles. We teste this effect in different conditions, and the results are summarized in Table\,\ref{tab:launching}.

\begin{table}[htbp]
\centering
\footnotesize
\setlength{\tabcolsep}{3pt}
\renewcommand{\arraystretch}{1.15}
\begin{tabular}{|>{\centering\arraybackslash}p{2.5cm}|>{\centering\arraybackslash}p{2.2cm}|c|>{\centering\arraybackslash}p{2.6cm}|>{\centering\arraybackslash}p{2.2cm}|c|>{\centering\arraybackslash}p{2.9cm}|}
\hline
\shortstack[c]{\textbf{Ring Voltage}\\\textbf{Amplitude [kV]}} &
\shortstack[c]{\textbf{Ring Voltage}\\\textbf{Signal Form}}   &
\shortstack[c]{\textbf{Frequency}\\\textbf{[Hz]}}             &
\shortstack[c]{\textbf{Glass Voltage}\\\textbf{Amplitude [kV]}} &
\shortstack[c]{\textbf{Glass Voltage}\\\textbf{Signal Form}}  &
\shortstack[c]{\textbf{Frequency}\\\textbf{[Hz]}}             &
\shortstack[c]{\textbf{Result}} \\
\hline
0.5--3.5 & AC & 1000 & 0   & grounded & -  & Null \\
\hline
0.5--3.5 & AC & 1000 & 0   & floating & -  & Null \\
\hline
0        & grounded &  -   & 0.5--3.5 & square  & 1   & Null \\
\hline
0        & floating &   -  & 0.5--3.5 & square  & 1   & Null \\
\hline
5.4      & AC & 200  & 3   & square   & 1   & Low flux launching \\
\hline
5.4      & AC & 200  & -4  & square   & 1   & High flux launching \\
\hline
5.4      & AC & 200  & -4  & DC       &  -   & High flux launching \\
\hline
5.4      & AC & 140  & 0   & grounded &   -  & High flux launching \\
\hline
4        & AC & 140  & 0   & grounded &   -  & Null \\
\hline
0        & grounded &  -  & 5.4 & AC & 200 & High flux launching \\
\hline
0        & floating &  -  & 5.4 & AC & 200 & High flux launching \\
\hline
0        & floating &  -   & 10  & DC &  -   & High flux launching \\
\hline
-10      & DC &   -   & 10  & DC &  -   & High flux launching \\
\hline
\end{tabular}
\caption{Summary of ND launching outcomes under different combinations of ring and glass voltages in vacuum. The table lists the applied amplitudes, signal forms, and frequencies for both glass and voltage, along with the resulting particle flux. “High flux launching” occurs only when piezo-induced vibrations from the ring are combined with an strong enough electric field, whereas “Null” and “Low flux launching” indicate conditions where the field strength isn't strong enough. The last combination of $20\rm\,kV$ voltage differences yields the highest flux of launched particles.}
\label{tab:launching}
\end{table}

\begin{figure}[htbp]

  \centering   \includegraphics[width=0.75\linewidth]{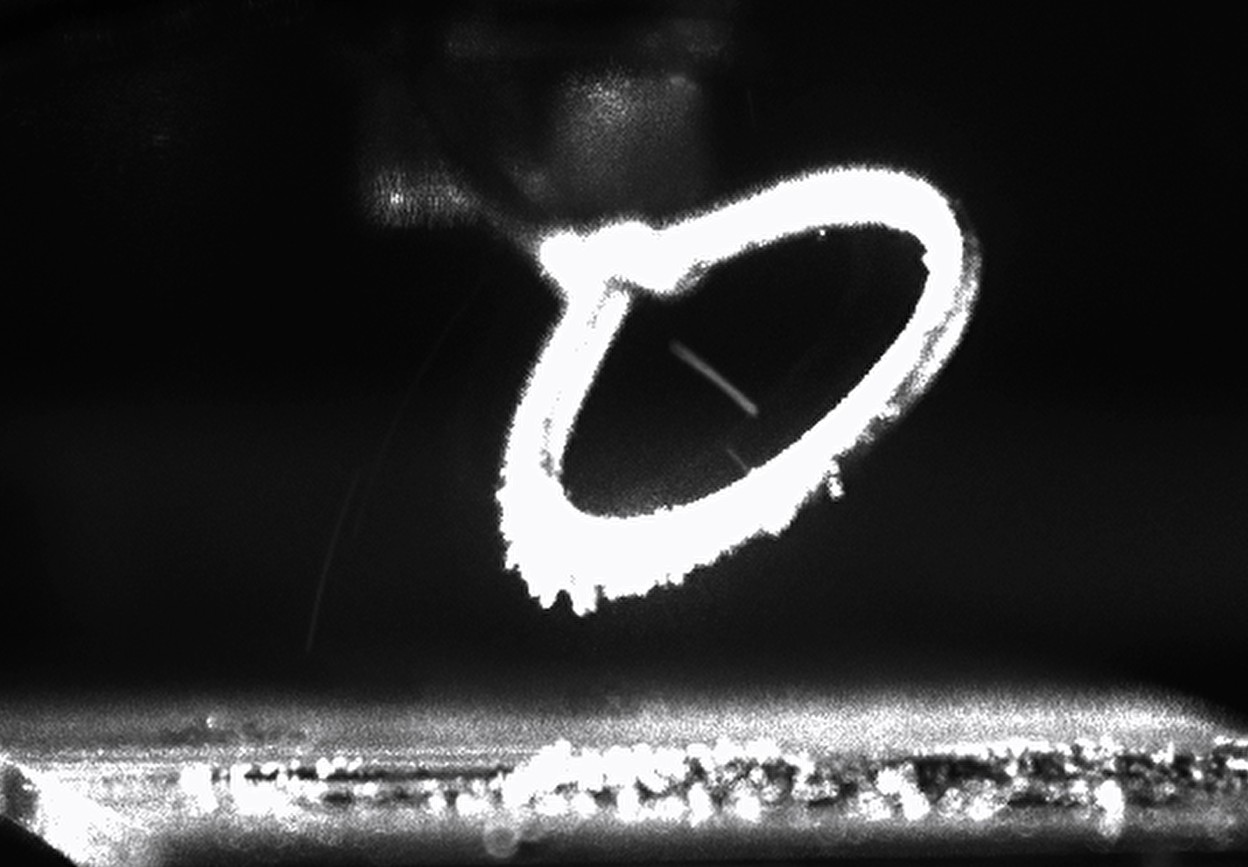}
  \caption{Electric launching of 75\,nm NDs at a pressure of \SI{e-5}{\milli\bar}  with a vibrating piezo and a voltage difference between the slide and ring of 20\,kV. Such high voltage differences were achieved using a manual Wimshurst Machine that was connected to each conductive part (ring and slide) and applying 10\,kV DC with opposite polarities. }
  \label{fig:electric_launching}
\end{figure}

Eventually, when applying a strong electric field, we observe the launching of charged particles. Particles are attracted upwards and some are stuck to the lower part of the ring, as shown in Fig.~\ref{fig:electric_launching}. In the successful experiments reported Table~\ref{tab:launching}, we report particles reaching their maximum height within 20–40\,ms. Since the experiments are conducted in vacuum conditions and our trap is conservative, successful trapping requires accurate timing of the trap. However, we face several challenges with this method: first, the maximum height durations are varied significantly; second, the launch itself is highly random in both timing and direction;  and third, we can't estimate the particle charge, and thus can't match the appropriate trap parameters to fulfill the stability conditions of the trap. Several works reported overcoming last challenge by trapping particles in air and then pumping the chamber to a lower pressure \cite{khodaee_dry_2022,Almuqhim_development_2019}. While this is possible, this approach does not fulfill the goal of this work, namely, clean loading in vacuum conditions.
As for launching NDs using piezo vibrations, we report an unsuccessful attempt to launch NDs upwards using the piezo vibrations alone at \SI{e-5}{\milli\bar}, while launching upwards is achieved by the combination of vibrations and electric field.

\section{Discussion}
While presently available NDs and ND sources are good enough for the first generation of ND SGI, we are working on improving the quality of the NDs \cite{givon2025fabricationnanodiamondssinglenv} as well as the source, while in parallel working to realize the first generation of a ND SGI with the available technology. 

Here we presented our survey of existing methods for loading nanoparticles into a Paul trap, as well as our own experimental investigation of piezo launching and electric launching, in air and vacuum. We find that piezo launching in vacuum conditions of $10^{-5}\,\rm mbar$ is highly inefficient. In air, piezo launching seems more effective, perhaps due to turbulence. Piezo launching in UHV also introduces significant heating into a chamber that, in the eventual ND SGI, will be cryogenic (in order to reduce decoherence due to blackbody radiation, and to increase the spin coherence time in the NDs). Electric launching was also found to be inefficient in vacuum conditions. However, electric launching in air was found to be highly efficient, both in terms of success rate and in terms of trapped particles vs the number of particles lost in the process. The high efficiency of this method makes it highly attractive for loading rare or expensive nanoparticles. While we do not fully understand the mechanisms that produce the electric launching and loading in air, it is clear from our experiments that breakdown of air is involved, and probably assists in charging the NDs, and maybe also applies forces on the charged NDs through the creation of plasma. To conclude, in all existing loading technologies, there are significant drawbacks that prevent us from using them for loading NDs into a Paul trap in vacuum conditions.

As high-quality NDs (size, shape, position of NV, coherence time) become more expensive, a high-efficiency source will become more and more advantageous. 
Hence, in the outlook, we present our design of a new ND source for Paul-trap levitation in UHV. This source is now being built and will soon be characterized. Preliminary results are shown.

\section{Outlook}\label{sec:linear_guide}

In this section, we present a design for a linear guide concentric with an embedded differential pumping tube. The goal of the apparatus is to allow loading of NDs in the open side of the guide (namely, in air) and to transport the NDs into a vacuum chamber. The idea for such a guide follows two similar techniques that have been previously realized. First, such electric guides are commonly used to direct ions \cite{douglas_linear_2005} in the field of mass spectrometry. Second, nanoparticles have been loaded through an optical guide of red-detuned light confined in a hollow-core fiber\,\cite{lindner_hollow-core_2024}. As explained previously, following Table \ref{tab:loading_trapping}, such an optical guide is inadequate for our needs as the ND absorbs light and heats up. Hence, we are constructing a dark guide similar to that being used for ions.

\begin{figure}[htbp]
  \centering
  \includegraphics[width=1\linewidth]{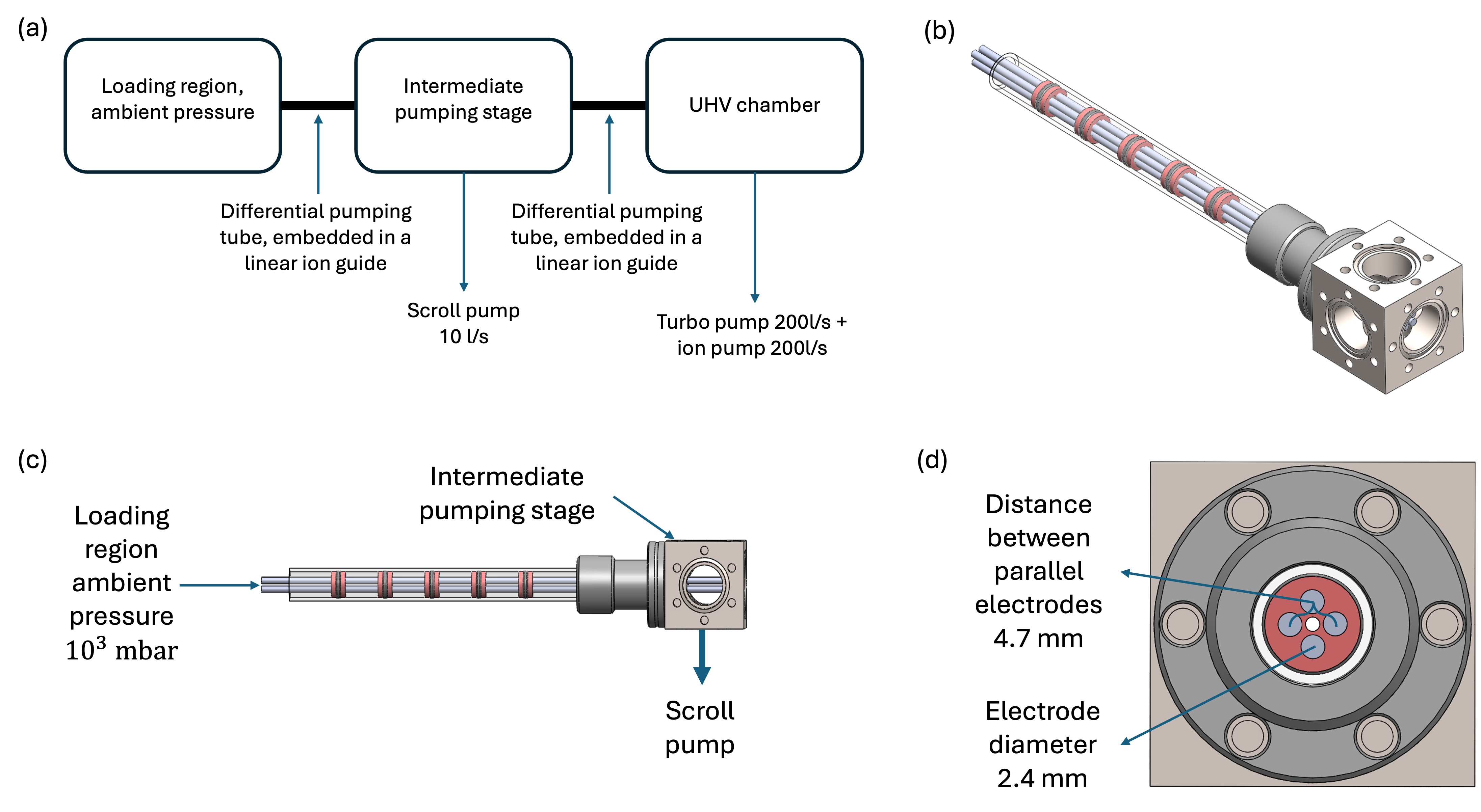}
  \caption{Schematics of a system for the efficient loading of a ND by transport via a linear guide. (a) General description of the system. From left to right, the loading is performed in ambient conditions (where we have already demonstrated high efficiency loading), from which the ND is transported into a second region (intermediate pumping stage) via a linear guide, embedded with a differential pumping tube. The intermediate pumping stage can also serve as storage for the NDs, or as a characterization stage. From there, the ND is transported into a UHV chamber via another differential pumping tube. (b) and (c): A 3D model of our design for the linear guide made from four stainless steel electrodes, held by five disks made of PEEK, in a glass tube, and connected to a CF16 cube, which serves as the intermediate pumping stage. (d) Front view of the 3D model, with the dimensions of the linear guide, where the diameter of the electrodes is 2.4\,mm, and the distance between the centers of the electrodes is 4.7\,mm. The center holes of the disks, which allow motion of the particles in the axis of the guide, act as a differential pumping tube and its dimensions are 6\,mm length and a diameter of 0.5-1.5\,mm, where these dimensions will be optimized after initial testing of the apparatus and full simulations of the gas and ND flow in the apparatus.}
  \label{fig:linear_guide}
\end{figure}

The linear guide consists of four parallel metallic electrodes, each with a diameter of 2.4\,mm, whose center is positioned 2.35\,mm from the symmetry axis of the guide. The geometry of the guide is inspired by the design given in \cite{Bullier_optomechanics_2020}. The electrodes are placed inside a glass tube, within which the linear guide is secured by several custom-made disks that hold the electrodes of the linear guide. The custom-designed disks are made from PEEK, each with five holes, four holes to hold the electrodes, and one hole in the center axis via which the ND can be transported. An O-ring on the circumference ensures a tight fit inside the tube. Namely, these disks hold the electrodes of the linear guide while the middle hole acts as a differential pumping tube, allowing a pressure difference between the loading region and the vacuum chamber, and allowing the NDs to pass through. The O-rings are tightly fitted to complete the barrier between the stages of the differential pumping tube
The whole construction (shown in Fig.\,\ref{fig:linear_guide}) is connected via an adapter to a CF16 cube, which can serve as a science chamber or a pumping stage to lower the pressure before the transport into a UHV chamber via a similar linear guide. 

In order for the ND to pass through the differential pumping tube, the amplitude of oscillation in the transverse direction should be smaller than the radius of the hole. We can calculate the amplitude using Eq.\,\ref{eq:secular_motion_amplitude}, and the following parameters: A ND with a diameter of 75\,nm, whose mass is given by
\begin{equation}
\begin{aligned}
2r &= 75\,\mathrm{nm}, \qquad
\rho &= 3500\,\mathrm{kg\,m^{-3}}, \qquad
m &= \frac{4\pi}{3}\,r^{3}\rho \approx 7.7\times 10^{-19}\,\mathrm{kg},
\end{aligned}
\end{equation}
in room temperature $T=300K$, Boltzmann constant $k_{B}=1.4\times 10^{-23}\,\mathrm{J\,K^{-1}}$, and a moderate trap frequency $\frac{\omega}{2\pi} =150\,\mathrm{Hz}$ \cite{Bullier_optomechanics_2020}. We then find an amplitude of:

\begin{equation}
\label{amplitude_of_secular_motion}
r_{0}=\sqrt{\frac{4k_{B}T}{m\omega^{2}}}=155\, \rm\mu m .
\end{equation}

We thus choose to make the differential pumping tube with a diameter in the range of $0.5 - 1.5 \,\rm mm$. We will test this range of pumping tube diameters in terms of particle transport and pressure difference to further refine the design.

\begin{figure}[htbp]
  \centering
  \begin{subfigure}[t]{0.48\linewidth}
    \centering
    \includegraphics[width=\linewidth]{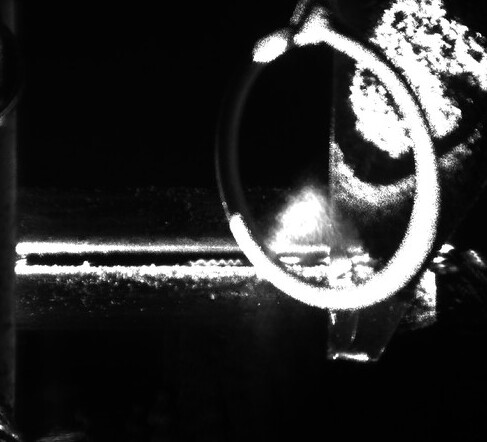}
  \end{subfigure}\hfill
  \begin{subfigure}[t]{0.48\linewidth}
    \centering
    \includegraphics[width=\linewidth]{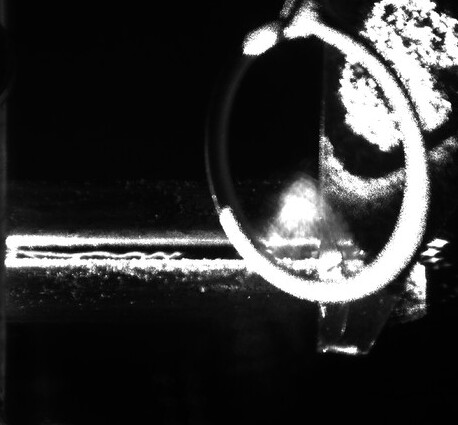}
  \end{subfigure}\hfill
  \caption{Trapping and guiding of NDs in the linear guide. NDs with a diameter of \(75\,\mathrm{nm}\) can be seen moving from right to left to right in the guide, exhibiting oscillations in the radial direction, resulting in a wavy trajectory. The particles are accelerated towards the vacuum chamber due to the pressure difference and resulting air flow. To load the particles, we use a rectangular, grounded, ITO-coated glass plate (visible on the right side of the guide), with a powder of NDs positioned on it. Above the glass, at a height of \(10\,\mathrm{mm}\), we place a ring electrode connected to an AC signal of \(4\,\mathrm{kV}\) amplitude at \(200\,\mathrm{Hz}\). The high voltage difference causes electrical breakdown of the air, which charges the particles and launches them using the same method described in Sec.~\ref{section:Results}. Some of the launched particles fall in the direction of the linear guide, which is driven by an AC signal of \(250\,\mathrm{V}\) amplitude at \(200\,\mathrm{Hz}\). The loading is done in ambient conditions, with the linear guide connected at its other end to a vacuum chamber at a pressure of \(46\,\mathrm{mbar}\). For this reason, the voltage on the guide is kept low enough to avoid electrical breakdown at low pressures. The pressure difference accelerates the guided particles toward the lower-pressure region. The transverse oscillations of the particles indicate that they are trapped in the plane normal to the guide axis.
}
  \label{fig:guiding_to_the_chamber}
\end{figure}

At this stage, we can already report preliminary results.
\begin{itemize}
    \item Successful loading and trapping of NDs in the linear guide, in ambient conditions, using the electrical launching method. A picture of such a trapped ND in the linear guide is shown in Fig.\,\ref{fig:ND_trapped_in_linear_guide}. This loading again confirms our suspicion that electrical breakdown of air is critical for the process, as the breakdown was even more visible in this setup, and correlated with the loading of particles into the trap.
    \begin{figure}[htbp]
  \centering
  \includegraphics[width=1\linewidth]{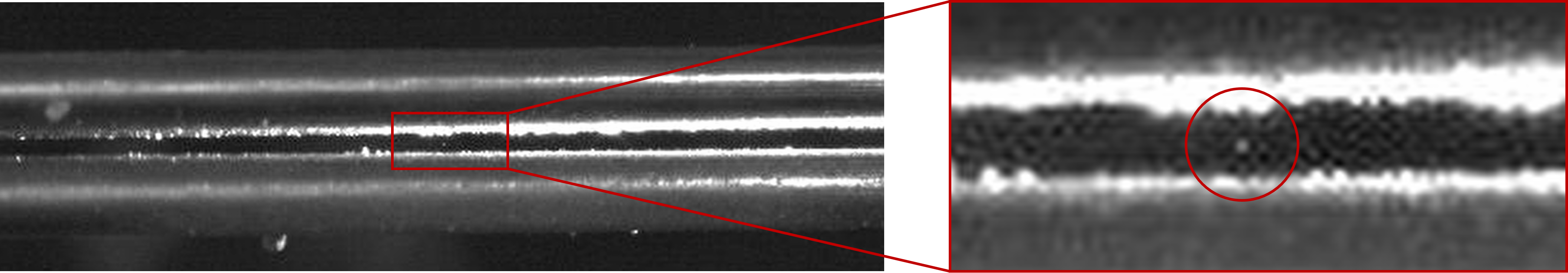}
  \caption{A trapped ND with a diameter of 75\,nm in the linear guide in ambient conditions. The particle was loaded from a conductive, grounded ITO-coated glass positioned beneath the guide. Two out of the four electrodes are visible in the picture. A green $532\rm\,nm$ laser shines in the middle of the guide, making the ND visible to the CCD camera.}
  \label{fig:ND_trapped_in_linear_guide}
\end{figure}
    \item
    A pressure drop from $10^3$\,mbar to $2.8 \times 10^{-2}$\,mbar was achieved using a two-stage pumping system. The first is pumped by a scroll pump and is separated from the atmosphere by a set of five disks with a hole of diameter 0.8\,mm, creating a differential pumping tube with an effective length of 3\,mm and diameter 0.8\,mm. The pressure in the first stage stabilized on a value of 46\,mbar. Between the first and second stage another set of four disks creates a differential pumping tube with an effective length of 2.4 mm and a diameter of 0.5\,mm. The second stage, in the main chamber, is pumped by a 90\,L/s turbo pump backed by a diaphragm pump, achieving a final pressure of $2.8 \times 10^{-2}$\,mbar. This represents a pressure difference of nearly five orders of magnitude, with the guide open to atmospheric pressure. When the guide is not in use, we can close the tube with a cap, and we observe a fast decrease in pressure to a value of $3\cdot10^{-4}$\,mbar. This vacuum system is presented in detail in Fig.~\ref{fig:linear_guide_in_real}.
\begin{figure}[htbp]
  \centering
  \includegraphics[width=1\linewidth]{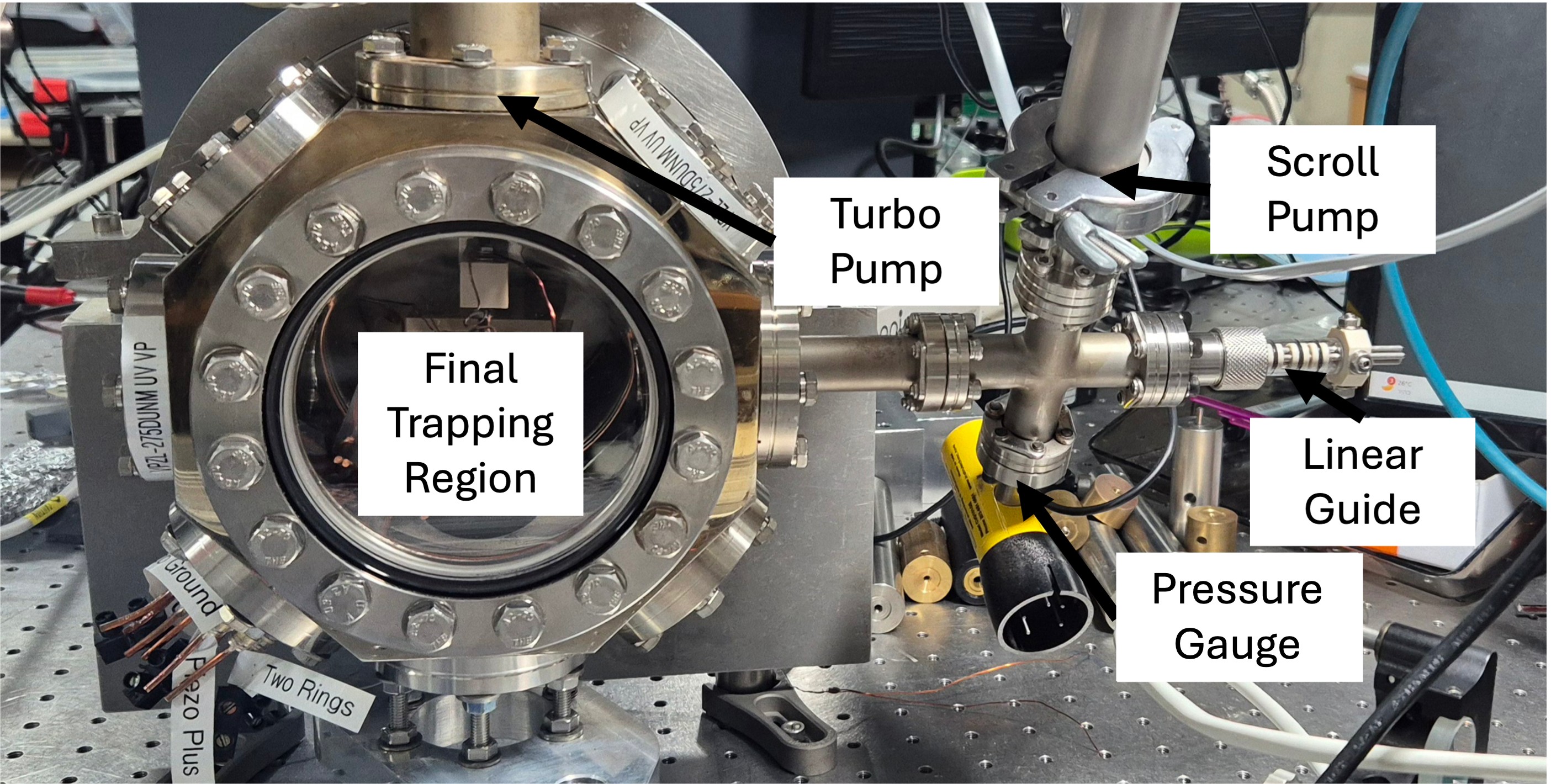}
  \caption{Preliminary vacuum chamber for linear guide loading. The linear guide can be seen connected to the right side of the chamber. To load particles into the guide, we place the high voltage ring around the end of the guide, and the grounded glass plate below it, as shown in Fig.~\ref{fig:guiding_to_the_chamber}. There is an intermediate pumping stage, which is pumped by a scroll pump, while the main chamber is pumped by a turbo pump, which is backed by a diaphragm pump. The particles loaded at the right end of the guide are accelerated towards the final trapping region due to the pressure difference. We measure a pressure of $2.8 \times 10^{-2}$\,mbar while the guide is open to air, and a fast decrease to a value of $3\cdot10^{-4}$\,mbar once we close the tube and guide with a cap.}
  \label{fig:linear_guide_in_real}
\end{figure}
    \item We observed successful guiding of NDs in the linear guide while the pumping is active. To load the guide, we use a ring Paul trap placed near the guide, with an ITO-coated glass slide between them. We turn on the guide, with an AC signal of $250\,\rm V$ amplitude and 200\,Hz. We then apply a high voltage signal on the ring, which creates a spray of charged particles, where some particles are emitted towards the guide. We observe particles trapped and guided towards the vacuum chamber due to the gas flow, as shown in Fig\,\ref{fig:guiding_to_the_chamber}.
\end{itemize}

To conclude, this design seems highly feasible for the following reasons: (1) It implements and combines well-known techniques, mainly a linear ion guide and differential pumping tube. While the two were not yet combined, each of them showed excellent results on its own in many applications. (2) The design is scalable and modular, in that it is possible to add intermediate pumping stages to reach the required level of vacuum, and that the guide and differential pumping tube can be custom-designed to accommodate the requirements of the experiment. (3) We have indications that it is possible to get high efficacy of loading NDs into the linear guide in ambient or near ambient pressure. (4) The preliminary results are positive, showing a pressure drop of nearly five orders of magnitude (with one intermediate pumping stage), as well as successful trapping and guiding of NDs in the linear guide.

\section*{Acknowledgments}
We thank the BGU Atom-Chip Group support team, especially Menachem Givon, Zina Binstock, Dmitrii Kapusta and Yaniv Bar-Haim for their support in building and maintaining the experiment. Funding: This work was funded by the Gordon and Betty Moore Foundation (\href{https://doi.org/10.37807/GBMF11936}{doi.org/10.37807/GBMF11936}), and the Simons Foundation (\href{https://www.simonsfoundation.org}{MP-TMPS-00005477}).

\end{document}